\documentclass[onecolumn]{emulateapj-rtx4}
\usepackage{graphicx}
\usepackage{epstopdf}
\usepackage{amsmath}

\providecommand{\abs}[1]{\lvert#1\rvert}

\usepackage{natbib} 

\begin{document}

\title{De-biased Populations of Kuiper Belt Objects from the Deep Ecliptic Survey}

\author{E. R. Adams\altaffilmark{1}, A. A. S. Gulbis\altaffilmark{2,3}, J. L. Elliot\altaffilmark{3,4}, S. D. Benecchi\altaffilmark{1,5}, M. W. Buie\altaffilmark{6}, D. E. Trilling\altaffilmark{7}, and L. H. Wasserman\altaffilmark{8}}

\altaffiltext{1}{Planetary Science Institute, 1700 East Fort Lowell, Suite 106, Tucson, AZ 85719}
\altaffiltext{2}{The Southern African Large Telescope and South African Astronomical Observatory, Cape Town, South Africa, 7935}
\altaffiltext{3}{Department of Earth, Atmospheric, and Planetary Sciences, Massachusetts Institute of Technology, 77 Massachusetts Ave., Cambridge, MA, 02139}
\altaffiltext{4}{Deceased}
\altaffiltext{5}{Carnegie Institution of Washington, Department of Terrestrial Magnetism, 5241 Broad Branch Road NW,  Washington, DC 20015-1305}
\altaffiltext{6}{Southwest Research Institute, 6220 Culebra Road, San Antonio, TX 78238, USA.}
\altaffiltext{7}{Department of Physics \& Astronomy, Northern Arizona University, S San Francisco St, Flagstaff, AZ 86011, USA}
\altaffiltext{8}{Lowell Observatory, 1400 W. Mars Hill Rd., Flagstaff, AZ 86001}

\begin{abstract}

The Deep Ecliptic Survey (DES) was a survey project that discovered hundreds of Kuiper Belt objects from 1998-2005.  Extensive follow-up observations of these bodies has yielded 304 objects with well-determined orbits and dynamical classifications into one of several categories:  Classical, Scattered, Centaur, or 16 mean-motion resonances with Neptune.  The DES search fields are well documented, enabling us to calculate the probability on each frame of detecting an object with its particular orbital parameters and absolute magnitude at a randomized point in its orbit.  The detection probabilities range from a maximum of 0.32 for the 3:2 resonant object $2002~GF_{32}$ to a minimum of $1.5*10^{-7}$ for the faint Scattered object $2001~FU_{185}$.  By grouping individual objects together by dynamical classes, we can estimate the distributions of four parameters that define each class: semi-major axis, eccentricity, inclination, and object size. The orbital element distributions ($a$, $e$, and $i$) were fit to the largest three classes (Classical, 3:2, and Scattered) using a maximum likelihood fit. Using the absolute magnitude ($H$-magnitude) as a proxy for the object size, we fit a power law to the number of objects vs. $H$ magnitude for 8 classes with at least 5 detected members (246 objects). The Classical objects are best fit with a power-law slope of $\alpha=1.02\pm0.01$ (observed from $5 \le H \le 7.2$). Six other dynamical classes (Scattered plus 5 resonances) have consistent  magnitude distributions \emph{slopes} with the Classicals, provided that the \emph{absolute number} of objects is scaled. Scattered objects are somewhat more numerous than Classical objects, while there are only a quarter as many 3:2 objects as Classicals. The exception to the power law relation is the Centaurs, which are non-resonant objects with perihelia closer than Neptune and therefore brighter and detectable at smaller sizes. Centaurs were observed from $7.5<H<11$, and that population is best fit by a power law with $\alpha = 0.42\pm0.02$. This is consistent with a knee in the H-distribution around $H=7.2$ as reported elsewhere \citep{Bernstein2004, Fraser2014}. Based on the Classical-derived magnitude distribution, the total number of objects ($H\le7$) in each class are: Classical ($2100\pm300$ objects), Scattered ($2800\pm 400$), 3:2 ($570\pm 80$), 2:1 ($400\pm 50$), 5:2 ($270\pm 40$), 7:4 ($69\pm 9$), 5:3 ($60\pm 8$). The independent estimate for the number of Centaurs in the same $H$ range is  $13\pm 5$. If instead all objects are divided by inclination into ``Hot'' and ``Cold'' populations, following \citet{Fraser2014}, we find that $\alpha_{Hot} =0.90\pm0.02$, while $\alpha_{Cold} = 1.32\pm0.02$, in good agreement with that work.

\end{abstract}


\section{Introduction}

Studies of the Kuiper Belt have progressed in the last two decades from finding individual objects \citep{Jewitt1993} to estimating the total number of objects in the outer solar system through observations and statistics \citep[e.g. this work;][]{Petit2011,Gladman2012}. The Kuiper Belt itself has been subdivided into many distinct dynamical classes of objects, including objects in mean-motion resonances with Neptune, high and low inclination populations \citep[Hot vs. Cold, e.g.][]{Morbidelli2008} and various groupings of scattered objects that have undergone dynamical interactions with Neptune. The relative numbers of objects in all of these populations offer one of the best direct observational constraints on different dynamical models of solar system formation and evolution.

A recent renaissance in modeling has resulted in a number of mostly successful attempts to reproduce the architecture of the solar system. Scattering and migration of the giant planets should leave directly observable signatures in the dynamical structure and numbers of small bodies remaining in the asteroid and Kuiper Belts and the outer satellites \citep{Levison2008, Morbidelli2009, Bottke2012}. Some models, such as the Nice Model \citep{Gomes2005, Morbidelli2005, Tsiganis2005} and the ``Grand Tack" \citep{Walsh2012}, invoke large-scale, abrupt motion by the giant planets.  Other models assume that giant planet migration proceeded more smoothly, and predict different ratios of objects captured into mean-motion resonances with Neptune \citep{Malhotra1993,Hahn2005}. 

Most models to date can successfully account for some -- but not all -- features of the Kuiper Belt, such as reproducing the Cold but not the Hot Classical populations (smooth migration), or leaving particular resonances over- or under-populated relative to the apparent populations (most models). Observational constraints can be used to distinguish between models and resolve fundamental questions, such as how Neptune migrated outward:  smoothly on a nearly circular orbit \citep[e.g.][]{Hahn2005}, through dynamic scattering with potential high eccentricity which was later damped down \citep[e.g.][]{Levison2008,Walsh2012}, or some other mechanism. This fundamental debate is evident in recent meetings, where one presentation concluded that chaotic capture models are more likely based on colors of resonant Kuiper Belt Objects, or KBOs \citep{Sheppard2012}, while another \citep{Noll2012DPS} claimed that the large binary fraction in the cold population favors smooth migration, since the mechanism of \citet{Levison2008} to push out the cold population would not preserve as many binary objects. However, it is also possible to have models with elements of both: a Nice-like initial migration that proceeds smoothly into a pre-existing cold population of objects that formed \emph{in situ} would not be ruled out by the binary fraction \citep{Parker2010b}.

Until recently, comparisons to observations have been disadvantaged because the known sample (over 1600 objects) is an inherently biased population, due to the difficulty of obtaining observations of faint, distant objects. What is needed is a de-biased population of objects, that is to say, a relatively large sample of objects discovered by a wide-field, all-sky survey that have had the sources of bias accounted for and removed. Two surveys with large discovery samples have recently begun to provide such results. The Canada-France Ecliptic Plane Survey (CFEPS), with a sample of 169 objects with high-quality orbits, reported de-biased population estimates in \citet{Petit2011} and \citet{Gladman2012}. 

Here we report the results of the Deep Ecliptic Survey (DES) \citep{Millis2002, Elliot2005}, the largest single survey to date, which has discovered about 500 objects with provisional designations by the Minor Planet Center (MPC). Of these objects, 304 have high-quality orbits suitable for dynamical classifications. The de-biased class populations presented in this paper offer a completely independent check on the CFEPS results, on both the absolute and relative numbers of objects, since the DES uses a different set of objects and discovery fields and a new approach to de-biasing. The DES sample also more than doubles the number of de-biased objects available, particularly in the small-number classes of high-order resonances and Scattered objects. Finally, we hope that the approach described in this paper will be useful for obtaining uniformly de-biased results of the expected thousands or tens of thousands of objects found by future large-scale surveys, such as the Large Synoptic Survey Telescope \citep[LSST, ][]{Ivezic2008} and Pan-STARRS \citep{Grav2011}. 

In Section~\ref{section:data} we describe the data and briefly review the choice of classification scheme. In Section~\ref{section:math} we present the analytical framework for calculating the detection probabilities. Section~\ref{section:numerical} contains the numerical implementation of the detection probabilities and the methods used to estimate the number of objects in each class. In Section~\ref{section:others} we compare our results to other surveys.  In Section~\ref{section:theory} we relate our observations to theories of planetary formation.

\section{Observations}
\label{section:data}

A full description of the DES can be found in \citet{Millis2002} and \citet{Elliot2005} (henceforth E05), but the relevant details are summarized here. The DES was an NOAO (National Optical Astronomical Observatory) survey. Two 4-meter telescopes (the Mayall at Kitt Peak in Arizona and the Blanco at Cerro Tololo in Chile) were used, with identical Mosaic cameras with 8 SITe CCD chips, each chip measuring 4096 by 2048 pixels. Each mosaic image covers about 0.35 deg$^2$ on the sky. Each field was within $\pm6.5^{\circ}$ of the ecliptic and was selected to have at least 35 USNO-B astrometric reference stars on each chip. The fields were also chosen to exclude stars brighter than magnitude $V=9.5$ \citep{Millis2002}. Typically two 300-sec exposures would be taken on the same night a few hours apart, with a third frame on another night during the same observing run. This resulted in orbital arcs of at least 24 hours (required for a designation by the MPC). Most of the fields imaged were distinct, but sometimes frames of the same field taken several years apart were counted as new search fields, since by that time a new set of objects would have moved into the field area. In 2005, the Deep Ecliptic Survey stopped surveying new fields, although recovery efforts to improve orbits and classifications of known objects have continued.

In order to obtain as uniform a sample as possible, for this work several selection criteria were applied to both the search fields and the objects discovered by the DES, as described below.

\subsection{Search frame selection} 

The discovery phase of the DES ended on 11 May 2005. A few KBOs were discovered during subsequent recovery efforts, but are not included in our sample since they were found under different search conditions. (For objects that were discovered during the main survey, we used the latest available orbital information to assign dynamical classes and calculate discovery probabilities.)

A total of 62,392 individual frames are in the DES database, corresponding to one of the eight CCD detectors on the Mosaic camera. Of these, 14,440 were not suitable for de-biasing because these frames were specifically targeted at the recovery of particular objects. Targeted recovery frames were not held to the same criteria for the number and brightness of stars, and have different statistical properties. An additional 1,312 frames with solar elongation angle less than $140^{\circ}$ were eliminated, because they have a different search efficiency than the bulk of the frames, which were taken near opposition. (Over twice as many objects were found on high solar elongation frames than on low-elongation frames.) 

The 46,640 individual frames left were then paired up with matching frames. Each field was observed at least three times: a pair of images taken on the same night a few hours apart, and one image on another night. Sometimes additional frames were taken, for instance if weather interrupted the original observations. Only the pair of frames on the same night was used to search for objects. There were 18,668 pairs of frames that met these criteria. The overlap fraction for each pair is quite high, with a median positional offset of 0.026\arcsec. Each frame is 1065\arcsec\ by 532\arcsec, and the largest mis-registration was less than 80\arcsec\ in RA and less than 90\arcsec\ in declination.

\subsection{Object selection}

As of 2012 August, a total of 913 KBOs were listed in the DES internal database. Some objects have been excluded from this analysis because they were found outside the main survey, during recovery or other follow-up observations (101 objects). We also exclude 8 objects found on low solar elongation fields ($<140^{\circ}$). Several additional objects, typically those that were lost shortly after discovery, are excluded because of missing information: objects lack discovery field coordinates (36 objects), discovery distances (104 objects), or discovery magnitudes (1 object). This leaves 663 objects discovered during the main DES survey. 

Of these objects, 478 objects had sufficient observations post-discovery to receive preliminary designations from the Minor Planet Center (72\%). Dynamical classification typically requires additional observations to achieve low enough errors, and 316 objects (48\%) have been classified (see Section~\ref{section:classification}). Almost all (304) of these classifications are considered to be secure (quality 3 as defined in E05). For our analyses, we used the 304 securely classified DES objects.

We note that there is a potential for recovery bias in the particular subset of objects that were both discovered and classified by the DES. One could imagine that objects in unusual orbits could be missed more frequently than objects that were closer to their predicted location when follow up observations were taken a few months later. However, we do not think that this recovery bias introduces a major problem for the current work for the following reasons. (1) The single biggest reason for failure to recover an object was insufficient telescope time to re-observe an object within 3 months of discovery, after which observations were typically not attempted; this affected objects in all orbits alike. It was more likely that an object was lost because we did not have enough clear weather nights than because we looked in the wrong place and did not find it. (2) When the estimated orbital parameters for the lost objects were compared to those that were ultimately designated and classified, \citet{Elliot2005} found that the only main difference was that the lost objects were fainter (and thus would be harder to recover on an average frame). In this work we find that the rate of recovery was twice as good for objects with $M_{disc} \le 23$ than for fainter objects, and we only do analyses using the better-recovered subset of brighter objects. Future work is planned to deal with the fainter objects, including the improved magnitude calibration, and we will also more fully explore the recovery bias at that time.

\begin{deluxetable}{c r r r r r r r }
\tablewidth{0pt}
\tablecaption{Objects per class}
\tablehead{
             & All                                         & ... also            	& ... also            & ... also		\\  
 Class  & $Q=3$\tablenotemark{a} &  $m_d \le 23$ 	& $H \le 7.5$   & $H \le 7.2$   	   
}
\startdata
 \text{1:1} & 1 & 1 & 1 & 0 \\
 \text{2:1} & 7 & 6 & 5 & 5 \\
 \text{3:1} & 1 & 1 & 0 & 0 \\
 \text{3:2} & 51 & 49 & 33 & 27 \\
 \text{4:1} & 1 & 1 & 1 & 1 \\
 \text{4:3} & 3 & 3 & 1 & 1 \\
 \text{5:2} & 7 & 6 & 6 & 3 \\
 \text{5:3} & 5 & 5 & 3 & 2 \\
 \text{5:4} & 3 & 3 & 3 & 2 \\
 \text{7:2} & 1 & 0 & 0 & 0 \\
 \text{7:3} & 3 & 3 & 2 & 0 \\
 \text{7:4} & 11 & 10 & 9 & 8 \\
 \text{9:4} & 3 & 3 & 3 & 3 \\
 \text{9:5} & 2 & 1 & 1 & 1 \\
 \text{11:6} & 1 & 1 & 1 & 1 \\
 \text{12:5} & 1 & 1 & 1 & 1 \\
 \text{Classical} & 144 & 130 & 129 & 122 \\
 \text{Centaur}\tablenotemark{b} & 8 & 7 & 1 & 0 \\
 \text{Scattered}\tablenotemark{c} & 41 & 33 & 24 & 22 \\
 \text{Far}\tablenotemark{d} & 10 & 9 & 7 & 6 \\
\hline
 \text{Total} & 304 & 273 & 231 & 205

 \enddata
\tablenotetext{a}{Secure orbital classifications (quality 3) only.} 
\tablenotetext{b}{The Centaur class in this analysis is restricted to dynamical Centaurs with $a \le a_{Nep}$.} 
\tablenotetext{c}{The Scattered class in this analysis is restricted to dynamical Centaurs, Scattered-Near or Scattered Extended objects with $a_{Nep} < a \le 80$~AU.} 
\tablenotetext{d}{Far objects are dynamical Centaurs, Scattered-Near or Scattered Extended objects  with $a > 80$~AU.} 
\label{table:det}
\end{deluxetable}

\subsection{Dynamical Classification}
\label{section:classification}

Over the years, several different schema for dynamically classifying KBOs have been proposed \citep[e.g.][]{Elliot2005, Lykawka2007, Gladman2008}. In this paper, we use the current DES classification scheme, a modified version of the scheme laid out in E05 which incorporates the Scattering concept from \citet{Gladman2008}. Since differences in the precise definitions of classes can affect the membership of objects and consequently the parameters derived for the class, we describe our classification scheme in full below.

To test for the dynamical class of an object, we use both the current observed orbital parameters as well as a 10 Myr integration of the orbit forward in time, and two additional 10 Myr integrations for the $\pm3~\sigma$ values for $a$, $e$, and $i$. An object is said to be securely classified (quality 3) only if all three integrations agree on the same dynamical class. (In quality 2 objects, the nominal classification only agrees with one of the $+3~\sigma$ or $-3~\sigma$ integrations, while in quality 1 it agrees with neither.) 

An object is tested for membership in each of the following dynamical classes, in order. (1) \emph{Resonant} objects occupy a mean-motion resonance with Neptune (resonances up to ninth order are identified from libration in the resonant angle). (2) \emph{Centaurs} have osculating perhelia that reach values less than the osculating semimajor axis of Neptune. (3) \emph{Scattered Near} objects have $a$ values that vary by more than an arbitrary amount, $\Delta a \ge 0.02$, over the 10 Myr integration (where $\Delta a = (a_{max}-a_{min})/a_{mean}$). This is along the same lines as the Scattering class of \citet{Gladman2008}, which uses $a_{max}-a_{min} > 1.5$~ AU. An excitation statistic, $s=\sqrt{e^2 + \sin(i)^2}$, is applied to the remaining objects, which are sent to two additional classes: (4) \emph{Scattered Extended} objects have $s \ge 0.25$, and (5) \emph{Classical} objects have $s<0.25$. Note that the exact criteria used can have an important impact on class membership: for instance, what might in other works be considered Hot Classical objects (low $e$, large $i$) with $i>15$ degrees would be classified as Scattered Extended under this scheme.

For the purposes of this paper, we have re-grouped a few objects in order to be more broadly consistent with the classes used by CFEPS and elsewhere. First, we restrict the population of Centaurs to only those with semi-major axes less than Neptune's (resulting in a population of smaller objects than found in any other class, due to the magnitude bias toward finding closer objects). Second, we group all of the dynamically excited objects with $a_{Nep}\le a \le80$~AU into a single \emph{Scattered} class, which contains objects from the Centaur, Scattered-Near, and Scattered-Extended classes. Ten objects with $a=82-739$~AU are not considered in this analysis, because they are too sparse in parameter space to adequately define a dynamical class; for a brief discussion of such rare objects see Section~\ref{section:rare}.


\section{Analytical framework}
\label{section:math}

We now consider how we can use the sample of KBOs discovered by the DES to learn about the entire population of KBOs.  Dynamical classes for KBOs have been defined with the underlying assumption that the members of each class have experienced a common formation process and common dynamical evolution.  Hence, we shall consider the objects in each dynamical class separately.  Within a dynamical class, we assume that the DES discoveries give us a sampling of the greater population of the objects, which can be characterized by four distribution functions: one each for the $H$-magnitude,  semimajor axis $a$, eccentricity $e$, and inclination $i$.

In Section~\ref{ss:prob}, we develop the equations to compute the probability of detection by the DES for each object, as a function of $H$, $a$, $e$, and $i$.  In Section~\ref{ss:distfunc}, we present the distribution functions that we assume are followed by the general population of the dynamical class for the same four parameters.  Finally, in Section~\ref{ss:maxlike} we develop a maximum likelihood model to fit the data.  From this fit, we determine, with error bars, the parameters of the four distribution functions and the number of objects in each dynamical class, within the range of orbital parameters probed by the DES discoveries.  We have chosen to apply the maximum-likelihood technique directly, rather than applying least-squares fitting to binned data, since the objects per bin are sparse in the four-dimensional $H$, $a$, $e$, and $i$-space \citep[see Chapter 10 of][]{Bevington2003}.

\subsection{Probabilities of Detection}
\label{ss:prob}

To compute the probability that an object would have been detected by the DES at some time during the survey, we extend the methods described in E05 to include observational biases introduced by an object's $H$, $a$,  $e$, and $i$ values.  We refer all inclinations to the Kuiper Belt plane (KBP), as derived in E05 ($i=1.74^{\circ} \pm0.23$, $\Omega = 99.2^{\circ} \pm 6.6$), which is consistent with the invariable plane of the solar system at the 1-$\sigma$ level.  Following the approach used in E05, we consider a set of $N_O$ objects ($j = 1, ..., N_O$) discovered in a set of $N_F$ search fields ($k = 1, ..., N_F$).  For the $j$th object discovered by the survey, we denote its $H$-magnitude, semimajor axis, eccentricity, and inclination by the symbols $H_j$, $a_j$, $e_j$ and $i_j$.  We assume that the $H$ magnitude of an object is not correlated with its orbital elements, and that angles that describe the orientation of the orbit in 3-dimensional space average out over time. (For resonant objects, this assumption is not quite true, and is addressed in a separate bias factor discussed below.)

Each of the $N_F$ fields of the survey data set is characterized by (i) the magnitude for which the detection efficiency has dropped to 1/2, which we denote by $m_{1/2,k}$, (ii) the Kuiper Belt latitude and orientation, or tilt, of the frame ($\beta_K$ and $\theta_K$), used to calculate which orbits pass through a field,  (iii) the Kuiper Belt longitude of the frame ($\lambda_K$), and (iiv) the solid angle, $\Omega_k$, subtended by the search field. Below we describe how biases related to all four of these characteristics have been removed.

(i) We assume that all search fields have the same maximum efficiency at bright magnitudes, $\epsilon_{max}$, and the same characteristic range, $\sigma_m$, over which the detection efficiency drops from $\epsilon_{max}$ to 0; as in E05, we have fixed the values of $\sigma_m=0.58$ and $\epsilon_{max}=1$.  For the $k$th search field, we employ the functional form used by \citet{Trujillo2001} and in E05 (their equation 19) for the detection efficiency as a function of magnitude, $\epsilon(m, m_{1/2,k})$:


\begin{equation}
\epsilon(m, m_{1/2,k}) = \frac{\epsilon_{max}}{2} \left[ 1+\tanh{  \left(  \frac{ m_{1/2,k} - m}{\sigma_m}  \right) } \right] .
\label{eqn1}
\end{equation}

Due to its elliptical orbit, the magnitude of the $j$th object varies from $m_{min,j}$ to $m_{max,j}$ (considering solely distance variations and ignoring physical changes).  Between these extremes, the relative time that the object is at magnitude $m$ is given by the probability distribution function $p_j(m)$ for the $j$th object.  Hence the likelihood, $\xi_{mag,j,k}$, for the $j$th object being discovered in the $k$th search field due solely to its magnitude (we shall consider other factors later) is given by

\begin{equation}
\xi_{mag,j,k} = \int_{m_{min,j}}^{m_{max,j}}  p_j(m) \epsilon(m, m_{1/2,k}) dm .
\label{eqn2}
\end{equation}

Our strategy for evaluating the integral in Equation~\ref{eqn2} involves transforming it from an integral over magnitude to an integral over heliocentric distance.  To do this we make several simplifying assumptions.  First we assume that all objects are discovered at opposition, and we set the Earth-Sun distance to 1 AU when computing the variation in the opposition magnitude of an object throughout its orbit.  We define $R_{d,j}$ as the heliocentric discovery distance for the $j$th object and $m_{d,j}$ as its discovery magnitude. We neglect photometric variability due to changing phase angle and/or rotational light curve, since the mean amplitude of variability is only 0.1 mag, with only 15\% of all KBOs varying by more than 0.15 mag \citep{Thirouin2010}. Then the magnitude of the $j$th object at a heliocentric distance $R$ is given by the equation,

\begin{equation}
m_j(R) = m_{d,j} + 5 \log_{10}(R/R_{d,j}) + 5 \log_{10} \left[ (R-1~AU) / (R_{d,j} - 1~AU) \right] .
\label{eqn3}
\end{equation}

For the $j$th object, the minimum and maximum heliocentric distances are $R_{min,j}$ and $R_{max,j}$ respectively:

\begin{equation}
R_{min,j} = a_j (1 - e_j)
\label{eqn4},
\end{equation}
\begin{equation}
R_{max,j} = a_j (1 + e_j)
\label{eqn5}.
\end{equation}

To find the amount of time that the $j$th object spends at a given heliocentric distance, $R$, it is possible to directly numerically integrate a two-body orbit and then calculate the amount of time spent in each distance bin. However, this approach is far too slow for an equation that is called frequently, and there are no good analytical approximations that work near perihelion and aphelion, where then probability of finding an object peaks sharply. We thus calculated a grid of numerical integrations using $e=0.001$ and $e=0.01-0.99$ ($\Delta_e=0.01$). For each eccentricity, we fit a piecewise function composed of five line segments ($f_1$ to $f_5$) at predefined break points( $p_1$ to $p_4$). The coordinates for the slopes and offsets and break-points of each line are stored for each eccentricity in a python executable file, and may be quickly called during numeric integrations. The function, which has been normalized over the allowed $R$ range, is scaled to the appropriate $a$ value when called. Tests indicate that the results are within $\pm15\%$ of the values obtained using numeric integration (for objects with $a<=80$~AU, which is the cutoff in this paper for the Scattered class).

\begin{equation}
p_j(R) = \begin{cases}
0								& R < R_{min,j} \\
f_1(e,R)					 		& R_{min,j} \le R \le p_1 \\
f_2(e,R)					 		& p_1 \le R \le p_2 \\
f_3(e,R)					 		& p_2 \le R \le p_3 \\
f_4(e,R)					 		& p_3 \le R \le p_4 \\
f_5(e,R)					 		& p_4 \le R \le R_{max,j} \\
0								 & R > R_{max,j} .
\end{cases}.
\label{eqn6}
\end{equation}

We are now in a position to rewrite Equation~\ref{eqn2} in terms of an integral over heliocentric distance.  Since our approximations reduce the apparent magnitude of an object with a given $H$ magnitude to only the variation that depends on its semi-major axis and eccentricity, we emphasize this by using the symbol $\xi_{j,k}(H,a,e)$ for the likelihood factor instead of $\zeta_{MAG,j,k}$ used in E05 (see the Glossary in their Appendix C).  Substituting $p_j(R)dR$ for $p_j(m)dm$ in the integral and changing the limits of integration from magnitudes to the corresponding heliocentric distances, we find

\begin{equation}
\xi_{j,k}(H,a,e) = \int_{R_{min,j}}^{R_{max,j}} p_j(R) \epsilon(m_j(R), m_{1/2,k}) dR.
\label{eqn7}
\end{equation}

Equation~\ref{eqn7} can be evaluated with substitutions from Equations~\ref{eqn1},~\ref{eqn3}, and~\ref{eqn6}, using Equations \ref{eqn4} and \ref{eqn5} to set the limits of integration.

(ii) We now consider the bias factor for inclination, $\xi_{j,k}(i)$, using the same approach as in the inclination distribution analysis of \citet{Gulbis2010}, hereafter G10. First we write an expression for the conditional probability of finding an object at latitude $\beta$ with an inclination $i$, which is Equation (9) of G10:

\begin{equation}
p(\beta | i) = \begin{cases} 
\frac{\cos \beta} {\pi \sqrt{\sin^2 i - sin^2 \beta}}				& \sin{i} > \abs{\sin \beta}  \\ 
 0 												& \sin{i} \le \abs{\sin \beta} ~\&~ i \ne \beta \ne 0\\
 1												& i = \beta =0
\end{cases}.
\label{eqn8}
\end{equation}
(Note that in G10 and this work all inclinations and latitudes are relative to the KBP.)

We define $\xi_{j,k}(i)$ as the likelihood (based on the inclination alone) of detecting the $j$th object in the $k$th search field, which we determine by integrating the conditional probability for the inclination, $i_j$, over the range of KBP coordinates covered by the $k$th search field: 

\begin{equation}
\xi_{j,k}(i) = \int \limits_{\beta_{min,k}}^{\beta_{max,k}}  \Delta \lambda (\beta'_k, \theta_k) p(\beta'_k | i_j) \,d\beta'_k,
\label{eqn9}
\end{equation}
where $\Delta \lambda$ is the range of longitudes at each latitude on the frame (a geometric parameter derived in G10, their Equations 11-14), and $\theta_K$ is the tilt of the frame with respect to the KBP.

(iii) Resonant objects, by definition, oscillate around the resonant angle with a certain amplitude. Observationally, this has the effect that they are observed preferentially at some longitude relative to Neptune but not at others. Using the libration amplitudes found in a 10 Myr integration of the orbit of each object, we can determine a \emph{longitude bias}, where the fields in the allowed longitude range, $\lambda_{j,min}  \le \lambda \le \lambda_{j,max}$, are upweighted and all other longitudes are set to zero (note that for non-resonant objets $\xi=1$): 

\begin{equation}
\xi_{long,j} = \begin{cases} 
360 / (\lambda_{j,max} - \lambda_{j,min})					& \lambda_{j,min}  \le \lambda \le \lambda_{j,max} \\
0												& \text{otherwise} 
\end{cases}.
\label{eqn10}
\end{equation}

(iv) A minor correction is included for the solid angle of each set of fields that is available for object searches. The effective search area for a pair of frames varies slightly due to mis-registration (typically a few arcseconds) and obscuration by other objects on the field. Defining $\Omega_s$ as the solid angle of a full CCD and $\Omega_k$ as the net solid angle for the kth CCD, the solid angle component of the likelihood factor (following E05) is

\begin{equation}
\xi_{ang,k} =\Omega_k / \Omega_s .
\label{eqn11}
\end{equation}

Neglecting any correlation among $a$, $e$, $i$, and $H$, the combined likelihood, $\zeta_{j,k}(H,a,e, i)$, for detecting the $j$th object in the $k$th search field is the product of the four separate likelihood components in Equations~\ref{eqn7},~\ref{eqn9},~\ref{eqn10}, and~\ref{eqn11}:

\begin{equation}
\zeta_{j,k}(H,a,e,i) = \xi_{j,k} (H,a,e)  ~ \xi_{j,k} (i) ~ \xi_{long,j} ~ \xi_{ang,k}.
\label{eqn12}
\end{equation}

If we had observed all the search fields simultaneously, then the probability for having detected the $j$th object for the $N_F$ search fields would be equal to the sum of the probabilities for each search field.  This would be the case for a single night, or even a single lunation.  However, we observe the search fields over a period of years, with the possibility of an object moving between search fields, so instead we take the efficiency of detecting the objects to be 1 minus the product of the likelihoods of having not detected it in each search field.  We refer to this quantity as the detection probability of the survey for the $j$th object, $q_{det}(H_j, a_j, e_j, i_j)$: 

\begin{equation}
 q_{det}(H_j, a_j, e_j, i_j) = 1 - \prod_{k=1}^{N_F} (1 - \zeta_{j,k} (H,a,e,i)).
 \label{eqn13}
 \end{equation}

From the probability of detecting each individual object and other assumptions, we can characterize the distribution of unbiased orbital parameters and estimate the numbers of objects in each of the dynamical classes, as described in the next section.

\subsection{Distribution Functions for $H$-Magnitude and Orbital Parameters}
\label{ss:distfunc}

For each class of objects, we want to determine the distribution functions for the parameters defining the class ($a$, $e$, $i$, and $H$). Here we present the functional forms used.

The simplest differential $H$ magnitude distribution function, $p_H(\alpha, H)$, is an exponential function with exponent $\alpha$. (We later present evidence for a double exponential, with two different power law indices following a break point, but since the break point lies near or outside of the range of the distribution functions for the classes we can fit, we only use a single exponent here.) All distribution functions are normalized over the full range of parameters considered, where $c_H$ is a normalization constant so that the integral of the distribution function between $H_{min}$ to $H_{max}$ is equal to one. Thus the form of the $H$ distribution is modeled using:

\begin{equation}
p_H(\alpha,H) = \begin{cases}
c_H 10^{\alpha H}		& H_{min} \le H \le H_{max}, \\
0					& H < H_{min} \mbox{ or } H_{max} < H,
\end{cases}
\label{eqn14}
\end{equation}

where

\begin{equation}
c_H = \frac{ \alpha \ln 10} {10^{\alpha H_{max}} - 10^{\alpha H_{min}} }.
\label{eqn15}
\end{equation}

Next we consider the differential distribution function for semimajor axes.  Between $a_{min}$ and $a_{max}$ for each class,  we model this distribution function, $p_a(a_o, \Delta a, a)$, as a Lorentzian with its central peak offset from zero by $a_o$ and full-width at half-maximum (FWHM) of $2\Delta_a$:

\begin{equation}
p_a(a_o, \Delta a, a) = \begin{cases}
\frac{c_a} {1+\left[(a-a_o)/ \Delta_a \right] ^2}		& a_{min} \le a \le a_{max}, \\
0										& a<a_{min} \mbox{ or } a_{max} < a,
\end{cases}
\label{eqn16}
\end{equation}

where the normalizing constant, $c_a$, is given by 

\begin{equation}
c_a = \frac{1}{ \Delta_a \left[ \arctan \left( (a_{max} - a_o) / \Delta_a \right) - \arctan \left( (a_{min} - a_o) / \Delta_a \right) \right]}.
\label{eqn17}
\end{equation}

The distribution of eccentricities is similarly modeled as a single Lorentzian with parameters  $e_o$, $\Delta_e$, and $c_e$, using analogous versions of Equations~\ref{eqn16} and~\ref{eqn17}. 

The inclination distribution is also represented as a single Lorentzian, using $i_o$, $\Delta_i$, and $c_i$. However, we note that other models have been used for the Classical distribution, notably a double Gaussian that represents two populations of objects, the ``core" and ``halo" populations \citep{Brown2001, Elliot2005, Gulbis2006b}, or alternately the ``hot'' and ``cold'' populations. To be specific, the first Gaussian component is the narrower one (core) and has a characteristic width (the standard deviation in the Gaussian expression) of $\Delta_{i1}$.  The second Gaussian component is broader (halo) with a characteristic width $\Delta_{i2} >\Delta_{i1} $.  We normalize each component separately, and $b$ is a number between zero and one, representing the fraction of objects in the narrower Gaussian component.  The inclination distribution function, $p_i(b, \Delta_{i1} , \Delta_{i2} , i)$, can be written as:

\begin{equation}
p_i(b,\Delta_{i1} , \Delta_{i2} , i) = \begin{cases}
\frac{b c_{i1}}{\sqrt{2\pi\Delta_{i1}^2} } \exp \frac{-i^2}{2\Delta_{i1}^2} + \frac{(1-b) c_{i2}}{\sqrt{2\pi\Delta_{i2}^2} } \exp \frac{-i^2}{2\Delta_{i2}^2}			& i_{min} \le i \le i_{max} \\
0				& i < i_{min} \mbox{ or } i_{max} < i
\end{cases}.
\label{eqn18}
\end{equation}

The normalizing constants, $c_{i1}$ and $c_{i2}$, are given by:

\begin{equation}
c_{i1} = \frac{1}{I(\Delta_{i1}, i_{max}) - I(\Delta_{i1}, i_{min}) }~\text{and}~
c_{i2} = \frac{1}{I(\Delta_{i2}, i_{max}) - I(\Delta_{i2}, i_{min}) }
\label{eqn19},
\end{equation}

where the integral, $I(\Delta, i)$ is given by the error function (erf):

\begin{equation}
I(\Delta, i) = \frac{1}{2} \mbox{ erf }( \frac{i}{\sqrt{2} \Delta} ).
\label{eqn20}
\end{equation}
Using the current definition of the Classical class, however, this double Gaussian model is not required (see Section~\ref{section:ml}).

\subsubsection{Range of normalization}
\label{ss:norm}

In order to include information from all frames, whether or not objects were detected on them, all distribution functions are normalized over the considered range of parameters.  The $e$ and $i$ values are normalized over the full range of parameter space ($e=0-0.95$ to avoid hyperbolic orbits, and $i=0-170^{\circ}$ to avoid numerical effects around 180 degrees; the effect of stopping the integrals just shy of the full range is insignificant, and no actual objects were found with those parameters). For the $a$ values, we used the minimum and maximum observed values in each class: the distribution of 3:2 objects is shaped by dynamics, and we do not expect it to continue outside of the observed range of semi-major axes ($a=39-40$~AU), while for the Classical and Scattered classes enough objects are observed far from the main peak of objects that a difference in a few AU on either side has very little effect on the normalization. The $H$ distributions, which are exponential functions, are normalized over the range of objects detected, and does not provide any information about brighter or fainter objects that the ones we found.

\subsection{Maximum Likelihood Estimation of Distribution Functions}
\label{ss:maxlike}

Using the functional forms defined in the previous section, we can now estimate the unbiased distribution of objects in $a$, $e$, $i$, and $H$, as well as the total number of objects per class. Both tasks are accomplished using a maximum likelihood (ML) method.

\subsubsection{Distribution Function Parameters}

For each dynamical class, we fit for the parameters that best fit the unbiased distribution functions. We do this using a likelihood function, $L$, which is given by a product over every discovered object, $j$, of the probability density functions, raised to the inverse power of the detection probability for that object:

\begin{equation}
L = \prod_{j=1}^{N_0} \left[ p_H(\alpha,H_j) p_a(a_o,\Delta_a,a_j) p_e(e_o,\Delta_e,e_j)  p_i(b, \Delta_{i1}, \Delta_{i2}, i_j)  \right]^{1/q_{det}(H_j,a_j,e_j,i_j)}
\label{eqn21}.
\end{equation}
Each class has a different set of parameters, and each distribution has been normalized over the range considered (see Section~\ref{ss:norm}) so that information from fields that did not detect objects is included in this fit.

Computationally, it is easier to maximize the natural logarithm of $L$, defined as $M$ :

\begin{equation}
M = \sum_{j=1}^{N_0} \frac{ \ln \left[ p_H(\alpha,H_j) p_a(a_o,\Delta_a,a_j) p_e(e_o,\Delta_e,e_j)  p_i(b, \Delta_{i1}, \Delta_{i2}, i_j)  \right]}{q_{det}(H_j,a_j,e_j,i_j)}
\label{eqn22}
\end{equation}

To calculate the errors, we assume that the number of objects in the sample is great enough so that the formal errors on the ML estimates approximately follow a Gaussian distribution, using the method of \citet{Crooke1999}. We define the matrix \textbf{X} as a $j \times n$ matrix, where there are $j$ objects and $n$ parameters ($\alpha, a_o, \Delta_a, e_o, \Delta_e, b, \Delta_{i1}$, and $\Delta_{i2}$ if a double-Gaussian inclination model). Each element $X_{j, n}$ is given by the expression:

\begin{equation}
X_{j, n} = \left. \frac{\delta M}{\delta x_n} \right|_{H=H_j, a=a_j, e=e_j, i=i_j } .
\label{eqn23}
\end{equation}

The correlation matrix, \textbf{C}, can be expressed in terms of \textbf{X}:

\begin{equation}
\boldsymbol{C} = [\boldsymbol{X}^T\boldsymbol{X}^{-1}] .
\label{eqn24}
\end{equation}

The variance of the parameter $x_n$ is given by the appropriate diagonal element, $C_{n,n}$, of the correlation matrix, and the standard deviation is found by taking the square root:

\begin{equation}
\sigma_n = \sqrt{C_{n,n}} .
\label{eqn25}
\end{equation}

\subsubsection{Inferred Number of Objects}

Once we have determined the parameters for all of the unbiased distribution functions, we can estimate the number of objects in the class, again using ML.  Within the range of $H$, $a$, $e$, and $i$ values that the ML distribution function parameters are valid, we assume that the total population for a dynamical class is well described by the distribution functions.  Given that assumption, we calculate the fraction of the total population detected by the DES, $f$, where $0 \leq f \leq 1$, by integrating the distribution functions over the detection probability of the range of object parameters:

\begin{equation}
f = \iiiint \limits^{H_{max}, a_{max}, e_{max}, i_{max}}_{H_{min}, a_{min}, e_{min}, i_{min}} q_{det}(H,a,e,i)~ p_H(\alpha, H)~ p_a(a_o,\Delta_a,a)~ p_e(e_o, \Delta_e, e)~ p_i(b, \Delta_{i1}, \Delta_{i2}) \,dH  \,da \,de \,di .
\label{eqn26}
\end{equation}

To turn the discovered fraction of objects into an estimated total number of actual objects, with error bars, we use the generalized binomial distribution, along with another ML function, where the observed number is $N_o$, the actual number of objects is $N_o'$, and $\Gamma$ is the gamma function:

\begin{equation}
L(N_o') = \frac{N_o'!}{N_o! (N_o' - N_o)!} f^{N_o} (1-f)^{N_o'-N_o} 
= \frac{ \Gamma(N_o' + 1) }{\Gamma(N_o+1) \Gamma(N_o' - N_o + 1)} f^{N_o} (1-f)^{N_o'-N_o} 
\label{eqn27}
\end{equation}
and thus
\begin{equation}
M(N_o') = \ln \left[  \frac{ \Gamma(N_o' + 1) }{\Gamma(N_o+1) \Gamma(N_o' - N_o + 1)}  \right] + N_o \ln f + {N_o'-N_o} \text{ln} (1-f) .
\label{eqn28}
\end{equation}

The number of objects, $N_o'$, is found by maximizing Equation~\ref{eqn28}, while the error is found by numerically calculating the total derivative:
\begin{equation}
\sigma(N_o') = 1/ \sqrt{\abs{\frac{d M(N_o')}{d N_o'}}} .
\label{eqn29}
\end{equation}

\section{Numerical implementation}
\label{section:numerical}

\subsection{Calculating detection probabilities}

Numerical implementation was done using Mathematica 8.0 and Python 2.7.3. Equation~\ref{eqn13} gives the probability of detecting an object with an orbit described by $a$, $e$, and $i$, an absolute $H$-magnitude, and which was discovered with magnitude $m_d$ at distance $R_d$. Values for $a$, $e$, $i$, and $H$ listed in the DES database in March 2012 were used in these calculations. 

Although 304 DES objects have the highest-quality orbital classification, we restrict our sample to the 273 objects discovered with a \emph{VR} magnitude of 23 or brighter. The nominal 50\% detection efficiency of the DES determined by E05 was $\text{VR}=22.5$, with $\text{VR}=23$ corresponding to an efficiency of 15\%, although the actual efficiency varies from frame to frame. The original photometric calibration of the DES (used in this and previous work) was based on the USNO-B1.0 catalog, which is however known to have magnitude uncertainties up to 0.5 mag. Recent recalibration of the DES \citep{Buie2011} has been completed but not yet reapplied to the survey results. Future work using the revised photometry will have much more reliable magnitudes, particularly for objects detected at the faint edge of the DES's range. Until the re-calibrated data is available,  we have calculated probabilities for all 304 objects, but restricted the de-biased class analysis to objects with $m_{d,\text{VR}} \le 23$. Table~\ref{table:det} shows the number of objects the DES found in each dynamical class, as well as the breakdown for objects with $m_d \le 23$ and additional $H$-magnitude constraints used in subsequent fits to single exponential functionss. Three non-resonant classes (Centaur, Classical, and Scattered) and sixteen resonances (five with $\ge 5$ objects) have been identified for the following analysis.

The detection probabilities (Equation~\ref{eqn13}) have been calculated for all objects with secure orbits, as shown in Column 2 in Table~\ref{table:prob}. The probabilities range from a maximum of 0.32 for the 3:2 resonant object $2002~GF_{32}$ to a minimum of $1.5*10^{-7}$ for the faint Scattered object $2001~FU_{185}$. (Similar low probability objects are discussed in Section~\ref{section:rare}.)

\begin{deluxetable}{c c  l l l l l l l l  }
\tablewidth{0pt}
\tablecaption{Detection probability for DES objects with secure classifications\tablenotemark{a}}
\tablehead{
Object & Class\tablenotemark{b}
 &  $a$& $e$ & $i$ & $H$ & $m_d$ & $R_d$ & Prob.\tablenotemark{c} & Fit\tablenotemark{d} \\
 &    & (AU) & & (deg) &  & (VR) & (AU) &
}
\startdata

 \text{2001QR322} & \text{1:1} & 30.38 & 0.0297 & 1.3438 & 7.42 & 21.1 & 29.654 & 0.26 & \text{} \\
\hline
 \text{2001FQ185} & \text{2:1} & 47.471 & 0.2258 & 2.2422 & 7.01 & 23.2 & 36.985 & 0.00016 & \text{} \\
 \text{2003FE128} & \text{2:1} & 47.714 & 0.2487 & 3.2774 & 6.17 & 21.5 & 36.019 & 0.019 & \text{} \\
 \text{2001UP18} & \text{2:1} & 48.008 & 0.0702 & 0.43497 & 5.71 & 22.5 & 50.271 & 0.061 & \text{} \\
 \text{2000QL251} & \text{2:1} & 48.043 & 0.2192 & 4.7652 & 6.54 & 22.2 & 38.211 & 0.011 & \text{} \\
...
\enddata
\tablenotetext{a}{This table is available in its entirety in a machine-readable form in the online journal. A portion is shown here for guidance regarding its form and content.}
\tablenotetext{b}{Centaurs with $a_{Nep}=30.1<a<80$~AU were grouped with Scattered objects for fits and plots throughout this paper. }
\tablenotetext{c}{Probability of detecting object with listed parameters and randomized ecliptic longitude.}
\tablenotetext{d}{Object used to derive fits. Centaur and Classical objects were used to derive CDF fits, and Classical, 3:2 and Scattered classes were used for maximum likelihood fits.}
\label{table:prob}
\end{deluxetable}

\subsection{Estimating size distribution}

Modelers work in physical sizes, while observers deal with apparent magnitudes, and properly comparing the results of the two requires some conversions and assumptions. The absolute, or $H$ magnitude, which is easily derived from the observations, is the preferred variable where both types of researchers can meet. Very few KBOs have actual measured diameters \citep[e.g.]{Elliot2010}, so observers would have to make assumptions about the object's physical properties, such as the albedo \citep[widely variable among Kuiper Belt objects,][]{Stansberry2008} to produce sizes. Using the apparent magnitude, as many observers and modelers have unfortunately done, is not recommended, because it conflates detection biases and leads to errors. We recommend that modelers report sizes in kilometers or else convert to $H$ magnitudes after noting the albedo assumed. 

With the data in a uniform framework it is now possible to compare whether the size (or rather, $H$ magnitude) distribution is the same across different dynamical classes. The implications of differences for the objects' collisional histories is discussed in Section~\ref{section:theory}. 

\subsubsection{Exponential fits from probability-weighted CDF}
\label{section:cdf}

\begin{figure}
\begin{centering}
\includegraphics*[scale=0.55]{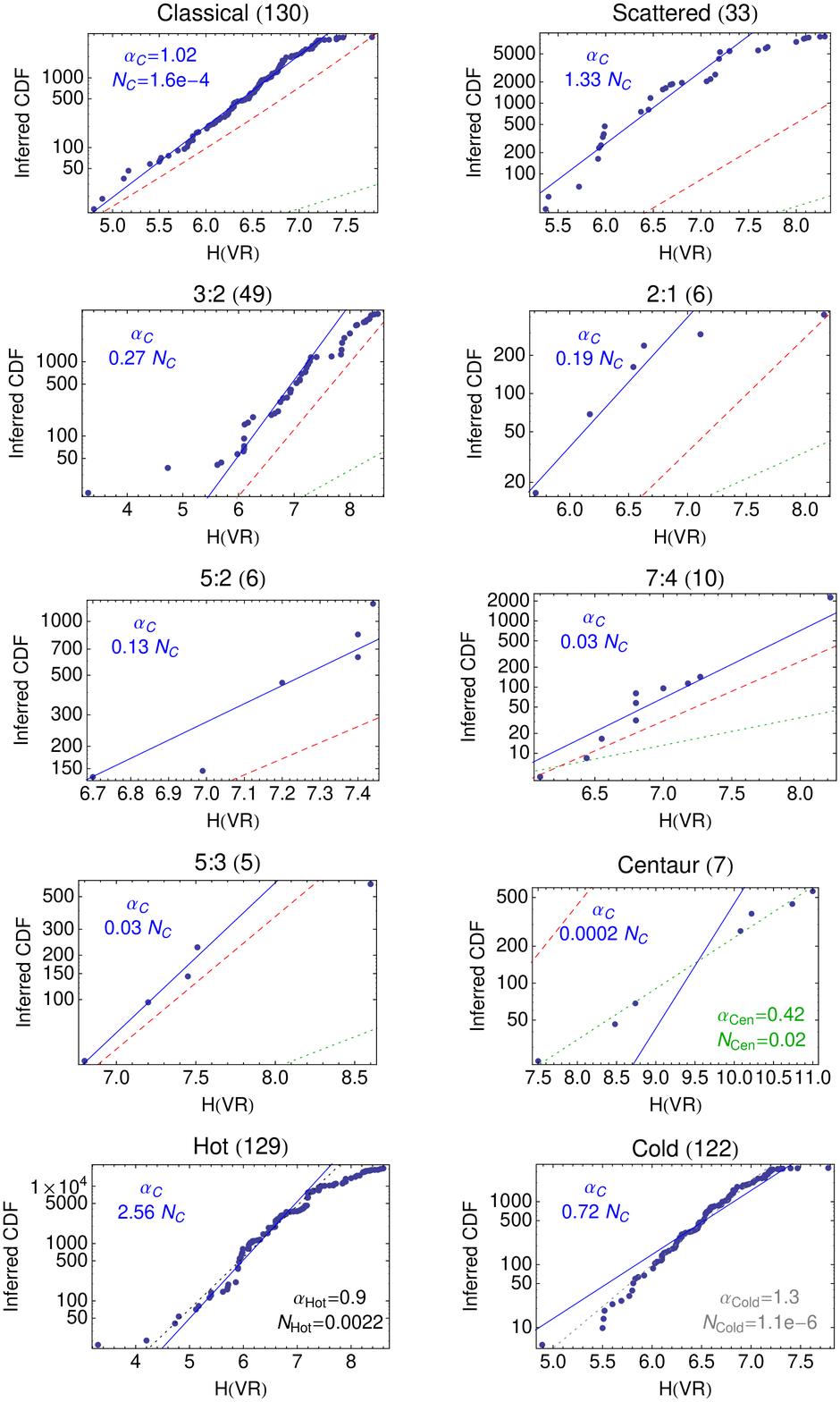}
\caption{Cumulative distribution function by $H$ magnitude. The dots show the probability-weighted number of each DES object (Equation~\ref{eqn30}). The fit for the Classical distribution ($4.8\le H \le 7.2$) has $\alpha_C = 1.02\pm0.01$. The blue line in each plot shows the Classical fit times a scale factor $N_X$ that varies by class. Three independent fits are shown: the Centaurs, which have a much shallower slope of $\alpha_{Cen} = 0.42\pm0.02$; and the ``Hot'' and ``Cold'' class of \citet{Fraser2014} that include objects of all dynamical classes (see Section~\ref{section:fraser}). The dashed red lines show the power laws for each class derived by CFEPS \citep{Gladman2012, Petit2011}, with values for $\alpha$ from 0.8 to 1.2. (Note that CFEPS did not provide a Centaur distribution, so the Scattered distribution is shown.)
 }
\label{fig:cdf}
\end{centering}
\end{figure}

Only classes with a relatively large number of objects can use the maximum-likelihood method above to find distribution functions for all parameters. However, even classes with as few as five objects can be fit for an $H$ magnitude distribution. A simple method to estimate the number of objects in a class is to convert the probability of detection into a predicted number of actual objects. A probability-weighted cumulative distribution of the estimated number of objects, $N$, brighter than a particular $H$ magnitude, where there are $N_o$ observed objects, is calculated by summing the inverse of the detection probability, $q_{det}$:
\begin{equation}
N (\le H) = \sum_{j=1}^{N_o} \frac{1}{q_{det,j}}.
\label{eqn30}
\end{equation}

For a given class $x$ and some normalization scale $C_x$, the CDF can be fit with an exponential function of the form
\begin{equation}
N_x (\le H) = C_x 10^{\alpha_x H}.
\label{eqn31}
\end{equation}

A single exponent $\alpha$ has been shown to break down when the range of $H$-magnitudes is large enough \citep{Jewitt1998,Gladman2001,Bernstein2004,Fuentes2008,Fraser2008,Shankman2013}. This is evident in a turnover to a shallower slope, and can be mathematically described by introducing two slopes, $\alpha_1$ brighter than a certain $H$-magnitude, referred to as the break, $H_b$, and  $\alpha_2$ for fainter objects: 
\begin{equation}
N_x  (\le H) =  \begin{cases} 
C_x ~10^{\alpha_1 H}											&  H \leq H_b \\ 
C_x ~10^{\alpha_1 H_b} ~ 10^{\alpha_2(H - H_b)}					&  H_{b} < H  
\end{cases}.
\label{eqn32}
\end{equation}

A plot of the probability-weighted CDF for each class is shown in Figure~\ref{fig:cdf}. We examined eight classes: Classical, Scattered, Centaur, 2:1, 3:2, 5:2, 5:3, and 7:4. We required that each class have $N_o\ge5$ objects with $m_d\le 23$, to avoid problems with sparse sampling at faint magnitudes (note that similar results are obtained using $m_d \le  23.5$). Similarly, a few objects in the Scattered class have such large semi-major axes (hundreds of AU) that they distort any population fit due to sparse sampling and much lower probabilities of discovery; all objects with $a>80$~AU are excluded from these analyses. 

\begin{figure}
\begin{centering}
\includegraphics*[scale=0.5]{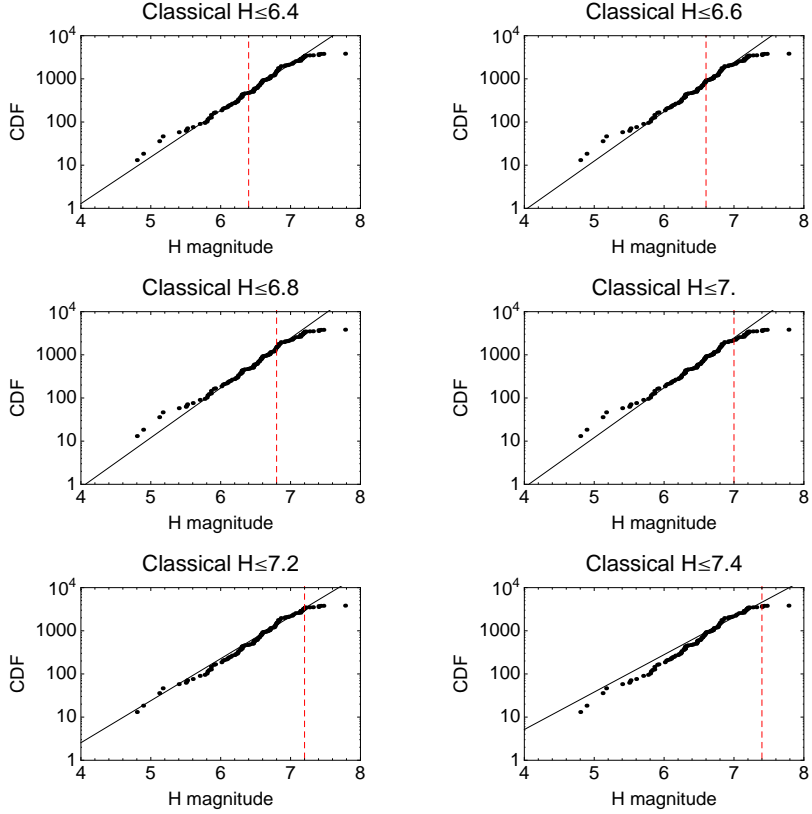}
\caption{Determining the break point, using the Classical distribution. All objects brighter than the break point, $H_b$ (dashed red line) are used to calculated the fit (solid line). The slope with the lowest error has $H_b=7.2$. 
 }
\label{fig:break}
\end{centering}
\end{figure}

A single exponential cannot explain the full distribution of objects. The location of the turnover to a shallower slope was estimated by examining the most populous class, the Classical objects. The single exponential fit with the lowest relative error has a break in slope at $H\leq 7.2$. A similar break is observed in the next two most populous classes, the 3:2 and Scattered objects, though the location of the break is not as well defined as in the Classical objects, since the 3:2 objects have more scatter at both high- and low-$H$ objects, and both the Scattered and 3:2 have a likely-coincidental lack of objects right around the putative break point. The slope for objects with $H<7.2$ is $\alpha = 1.02\pm0.01$ from the Classical data; fits to several break points from 6.4 to 7.4 are shown in Figure~\ref{fig:break}.

\begin{deluxetable}{ c c r r }
\tablewidth{0pt}
\tablecaption{CDF fits to power law slope}
\tablehead{
 \text{Parameter} & \text{Value}			& \text{Fitted range}			& \text{Derived from}
}
\startdata
$\alpha_1$	& $1.02\pm0.01$	&	$4.8\le H\le7.2$		& Classical CDF \\
$\alpha_2$	& $0.42\pm0.02$	&	$7.5\le H\le11.$		& Centaur  CDF \\
\hline
\text{2:1}	&	$0.93\pm0.17$	&	$5.7\le H\le7.1$	&	\text{5 objects}	\\
\text{3:2}	&	$0.84\pm0.03$	&	$5.6\le H\le7.4$	&	\text{31 objects}	\\
\text{5:2}	&	$1.29\pm0.23$	&	$6.7\le H\le7.4$	&	\text{6 objects}	\\
\text{5:3}	&	$1.00\pm0.12$	&	$6.8\le H\le7.5$	&	\text{4 objects}	\\
\text{7:4}	&	$1.29\pm0.09$	&	$6.1\le H\le8.2$	&	\text{10 objects}	\\
\text{Classical}	&	$1.02\pm0.01$	&	$4.8\le H\le7.2$	&	\text{122 objects}	\\
\text{Scattered}	&	$1.05\pm0.06$	&	$5.4\le H\le7.3$	&	\text{23 objects}	\\
\text{Centaur}	&	$0.42\pm0.02$	&	$7.5\le H\le11.$	&	\text{7 objects}	\\
\text{Hot}	&	$0.90\pm0.02$	&	$4.2\le H\le 7.2$	&	\text{84 objects}	\\
\text{Cold}	&	$1.32\pm0.02$	&	$4.9\le H\le 7.2$	&	\text{114 objects}	
\enddata
\label{table:cdffit}
\end{deluxetable}

We note that break points are commonly seen at the faint edge of observational data, although the observed slope after the break varies by group. \citet{Fuentes2008} observed objects spanning the break, which they placed at $R=24.3$ with a slope of $\alpha_1=0.7^{+0.2}_{-0.1}$ before and $\alpha_2=0.3^{+0.2}_{-0.3}$ after. This corresponds to a diameter of $D=118 (p/0.05)^{-0.5}$ km (where albedo $p = 0.05$), or an $H$-magnitude of about 8.5. The break seen in the DES exponential for Classical objects, at $H=7.2$, could thus perhaps be a limiting magnitude issue, if the DES survey efficiency was being mis-estimated at the faintest, most difficult to detect magnitudes. To test the hypothesis that the limiting magnitude was causing an artificial turnover, we tightened our discovery magnitude threshold from $\text{VR}=23$ to $\text{VR}=22.5$. The same exponent was found, but the turnover in the Classical population appeared at $H\le6.7$ instead. 

However, recent work by \citet{Fraser2014} also finds break points at a similarly low $H$, with their Cold population breaking at $H_{b,Cold}=6.9^{+0.1}_{0.2}$ and the Hot population at $H_{b,Hot}=7.7^{+1.0}_{0.5}$ in $r'$. Since their method is different (see Section~\ref{section:fraser} and their data is independent from what is presented here, it lends credence to the break point we have identified around $H_b = 7.2$ being real.

\begin{deluxetable}{ c c c c  c c c c  | c c }
\tablewidth{0pt}
\tablecaption{Estimated number of objects less than $H$}
\tablehead{
 \text{$\le $H} & \text{Classical} & \text{Scattered} & \text{3:2} & \text{2:1} & \text{5:2} &  \text{7:4} & \text{5:3} & \text{Centaur} 
}
\startdata
\multicolumn{8}{c}{(1) $\alpha_1 = 1.02$, $H_b=7.2$, $\alpha_2=0.42$, $N_C=1.6*10^{-4}$ }  & (2)	$\alpha_{Cen}=0.42$\\
\multicolumn{8}{c}{ }  & ~~ $N_{Cen} = 0.02$\\
\\
4 & $2\pm 0$ & $2\pm 1$ & $0\pm 1$ & $0\pm 0$ & $0\pm 0$ & $0\pm 0$ & $0\pm 0$ & $1\pm 0$ & \\
5 & $19\pm 2$ & $26\pm 2$ & $5\pm 1$ & $4\pm 0$ & $2\pm 1$ & $1\pm 0$ & $1\pm 0$ & $2\pm 0$ & \\
6 & $200\pm 30$ & $270\pm 30$ & $54\pm 7$ & $38\pm 5$ & $26\pm 3$ & $7\pm 0$ & $6\pm 0$ & $5\pm 2$ & \\
7 & $2100\pm 300$ & $2800\pm 400$ & $570\pm 80$ & $400\pm 50$ & $270\pm 40$ & $69\pm 9$ & $60\pm 8$ & $13\pm 5$ & \\
7.2 & $3400\pm 500$ & $4500\pm 600$ & $910\pm 90$ & $640\pm 90$ & $440\pm 60$ & $110\pm 20$ & $95\pm 15$ & $16\pm 6$ & \\
7.5 & $4500\pm 700$ & $6000\pm 900$ & $1200\pm 200$ & $850\pm 130$ & $580\pm 100$ & $150\pm 20$ & $130\pm 20$ & $21\pm 9$ & \\
8 & $7300\pm 1300$ & $9600\pm 1400$ & $2000\pm 300$ & $1400\pm 200$ & $940\pm 160$ & $240\pm 40$ & $200\pm 40$ & $35\pm 15$ & \\
9 & $19000\pm 5000$ & $25000\pm 6000$ & $5100\pm 1200$ & $3600\pm 800$ & $2500\pm 600$ & $620\pm 150$ & $530\pm 130$ & $90\pm 50$ & \\
10 & $49000\pm 15000$ & $66000\pm 19000$ & $13000\pm 4000$ & $9300\pm 2700$ & $6400\pm 1900$ & $1600\pm 500$ & $1400\pm 400$ & $240\pm 130$ & \\
\hline
  scale & 1.0 & 1.33 & 0.27 & 0.19 & 0.13 & 0.03 & 0.03 & NA
 \enddata
\label{table:numobj}
\end{deluxetable}

In addition, we have one population that clearly samples the size distribution \emph{after} the break in slope: the Centaurs. With 7 objects spanning a large range of $H$-magnitudes after the purported break ($7.5<H<11$), the Centaurs are well fit by a single, shallower exponential, with $\alpha = 0.42\pm0.2$.  (We note that similar slopes can be fit to the few Classical and 3:2 objects with $H>7.2$, though the exact slopes are strongly subject to the exact choice of the artificial break point and have a much smaller lever arm spanning only 0.5-1 magnitudes.)

Our proposed model for the size distribution of the Kuiper Belt is to assume that brighter objects follow the Classical $H$ magnitude distribution for $5\le H\le7.2$, and then shifts to a shallower shope for smaller objects as derived from the Centaur distribution. (There are hints in the 3:2 distribution that brighter objects may follow a shallower slope, but we have very few objects with $H<5$, and none are Classical objects.) Combining the bright slope fit to Classical objects with $H\le7.2$ with the faint slope fit to Centaurs with $H\ge7.5$, we can construct a double exponent: $\alpha_1=1.02\pm0.01$, $\alpha_2 = 0.42\pm0.02$, and $H_b=7.2$ (see Table~\ref{table:cdffit}). Note that the exact location of the break point may be an artifact of the limiting magnitudes of our fields, as very few objects (especially Classical) were found  fainter than $M>23$, and so our CDFs suffer from incompleteness.

We check the compatibility of each of the eight classes of KBOs with this double exponent using a Kolmogorov-Smirnov (KS) test, with one additional variable: although the values for $\alpha_1$, $\alpha_2$, and $H_b$ are kept the same, the absolute number of objects is allowed to vary. The reasoning behind this choice is that objects in a resonance may have the same overall size distribution, but due to varying efficiencies of resonance capture may differ in absolute abundance. At the 5\% level, all eight dynamical classes are compatible with this double exponent. The scaling constants required in the region following $\alpha_1$, relative to the number of Classical objects, $N_C = 1.6 * 10^{-4}$, are as follows: $N_{Scattered} = 1.33~ N_C$, $N_{3:2} = 0.27~ N_C$, $N_{2:1}=0.19~ N_C$, $N_{5:2}=0.13~ N_C$, $N_{7:4} = 0.03~ N_C$ and $N_{5:3} = 0.03~ N_C$. The Centaurs, which follow $\alpha_2$, have a different scale factor, $N_{Centaur} = 0.02$. In addition, the Hot and Cold populations are also consistent on a KS test with the Classical distribution, with $N_{Hot} = 2.56~ N_C$ and $N_{Cold} = 0.72~ N_C$. Visually, this agreement can be seen in Figure~\ref{fig:cdf}, which shows the scaled Classical fit with the data for each class.

We can estimate the total number of objects in each class less than a particular $H$-magnitude. Table~\ref{table:numobj} shows the estimated total number of objects in each class from $H=4-9$, using two models: (1) a scaled broken exponential with $H_b=7.2$ for all dynamic classes \emph{except} the Centaurs, and (2) a single exponential for the Centaurs. The errors are calculated by taking the difference between the nominal number and the numbers derived from the $\pm1~\sigma$ errors. Note that the Centaurs are estimated using their own independent fit and errors.

\subsubsection{Estimating class distribution parameters}
\label{section:ml}

For those objects with more members, we can more fully characterize the class properties. This approach uses a maximum-likelihood (ML) method to model the distribution functions for the parameters ($a$, $e$, $i$, and $H$) that define the class. These distributions are then used to calculate the detected fraction, $f$, from Equation~\ref{eqn26}, to find the total number of objects within our parameter range. A downside to maximum likelihood is that more objects are needed in a class to full characterize the four-dimensional parameter space. Only the Classical, Scattered, and 3:2 classes have sufficient objects to attempt a ML fit, and the Scattered fit is marginal. This is why we recommend using the values derived from the CDFs in Section~\ref{section:cdf} for comparison between many classes of objects.

We used somewhat expanded subsets of objects for the Scattered and 3:2 classes compared to the CDF analysis. For the ML analyses, a well-sampled parameter space is important, and we examined cutoffs for $H$-magnitude from $H=6.5-8.5$. For Classical objects, setting the break point between $H=6.7-7.2$ does not substantially change the fitted value for $\alpha$ (e.g., $\alpha_{H\le6.7}=1.05\pm0.12$), but including fainter objects does ($\alpha_{H\le7.5}=0.60\pm0.07$). The distributions for $a$, $e$, and $i$, however, are resilient to the choice of $H_b$.  Choosing a lower (brighter) cutoff means decreasing the sample size, and since there were few enough objects to start with, the cutoff was set at $H=7.5$ for the Scattered and 3:2 classes, while Classical objects use the previously determined CDF cutoff of $H=7.2$. 

\begin{figure}
\begin{centering}
\includegraphics*[width=0.7\textwidth]{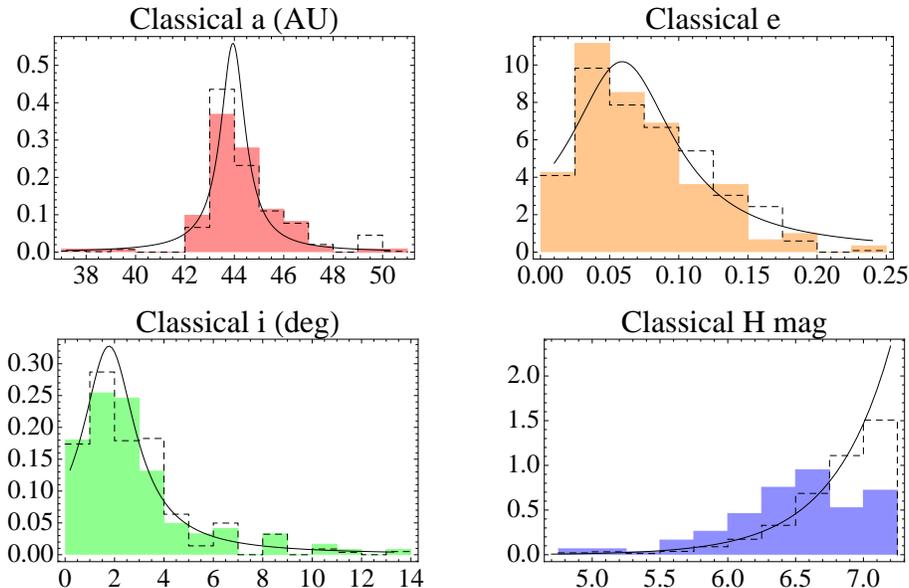}
\caption{Maximum likelihood fit to 122 Classical objects with $m_d \le 23$, and $H \le 7.2$. The filled histogram shows the observed distribution, while the dashed histogram shows the amount by which the distribution is adjusted to account for discovery probabilities. The black lines show the best maximum likelihood fit to each parameter.  
 }
\label{fig:mlclassical}
\end{centering}
\end{figure}

For all three classes we chose to model the $a$,  $e$, and $i$ distributions using a single Lorentzian (Equation~\ref{eqn16}),  following the preference for Lorentzians over Gaussians of E05 and G10. The Classical inclination distribution is also well fit by a double Gaussian (as in G10), but since the resulting distributions are not meaningfully different, the simpler model is preferred in this work. (Note: whether a single or double Gaussian is preferred depends critically on the exact classification scheme used. Four high-inclination objects that were deemed Classical objects in \citet{Gulbis2010} are classified as Scattered-Extended objects in this work, and that appears to explain why the double-Gaussian fit is not required here. Choice of classification scheme matters, particularly when comparing results!) The distributions and histograms of the biased and de-biased detections are shown in Figure~\ref{fig:mlclassical} (Classical), Figure~\ref{fig:ml3to2} (3:2), and Figure~\ref{fig:mlscattered} (Scattered).

We list the best fit values for the Classical, 3:2, and Scattered distributions in Table~\ref{table:mlfit}. The values for $\alpha$ agree within their errors with the slopes found for the Classical objects in the CDF analysis. The distributions were normalized over the observed range of parameters for $a$ and $H$\footnote{Minimum/maximum values for Classical objects: $37.2 \le a \le 50.3$ AU and  $4.7 \le H \le 7.2$; Scattered objects: $32.5 \le a \le 68.1$ AU and $5.3 \le H \le 7.3$; 3:2 objects: $39 \le a \le 40$ AU and  $4.7 \le H \le 7.4$; }, and from $0 \le e \le 0.95$ and $0 \le i \le 170^{\circ}$ (the full range has been truncated to avoid numerical errors near boundaries). We note that the spread in the eccentricity distribution for the Scattered objects is artificially narrow due to a small number of low-probability objects at $e=0.18$ (an effect that would presumably go away with more objects).

\begin{deluxetable}{ c c c c }
\tablewidth{0pt}
\tablecaption{Maximum likelihood fitted distribution parameters\tablenotemark{a}}
\tablehead{
 \text{Param} & \text{Classical}\tablenotemark{b} & \text{3:2}\tablenotemark{c} & \text{Scattered}\tablenotemark{d}
}
\startdata
$\alpha$  &  $1.01\pm0.07$  &  $0.95\pm0.16$  &  $1.04\pm0.19$\\
$a_o$  &  $43.94\pm0.09$  &  $39.36\pm0.05$  &  $45.87\pm0.24$\\
$a_s$  &  $0.61\pm0.09$  &  $0.2\pm0.07$  &  $0.67\pm0.28$\\
$e_o$  &  $0.06\pm0.01$  &  $0.24\pm0.01$  &  $0.18\pm0.01$\\
$e_s$  &  $0.05\pm0.01$  &  $0.03\pm0.01$  &  $0.01\pm0.004$\\
$i_o$  &  $1.78\pm0.1612$  &  $14.37\pm1.52$  &  $21.99\pm0.77$\\
$i_s$  &  $1.31\pm0.24$  &  $4.14\pm1.68$  &  $1.69\pm0.56$\\
$H \le 7.$  &  (2800)  & (840)  &  (3100)\\
$H \le 7.2$  &  $4500\pm400$  &  (1300) &  (5000)\\
$H \le 7.5$  &  (9000)  &  $2500\pm450$  &  $10000\pm2100$
 \enddata
\tablenotetext{a}{Parentheses indicate derived quantities.}
\tablenotetext{b}{Objects fit: quality 3, $m_d \le 23$, and $H \le 7.2$.}
\tablenotetext{c}{Objects fit: quality 3, $m_d \le 23$, and $5 \le H \le 7.5$.}
\tablenotetext{d}{Objects fit: quality 3, $m_d \le 23$, and $H \le 7.5$.}
\label{table:mlfit}
\end{deluxetable}

Using the calculated distribution functions, we evaluate Equation~\ref{eqn26} along a grid of 4-7 points each in $a$-$e$-$i$-$H$ space to find the detected fraction, $f$. The grid was constructed assuming that each grid object was discovered at its median distance from the sun (i.e., half its time is spent closer to the sun, and half its time is further out) and with a hypothetical discovery magnitude at that distance calculated using:

\begin{equation}
m_d = H + 5 \log_{10}(R_d / 1) + 5 \log_{10}(R_d - 1/ 1) .
\label{eqn33}
\end{equation}

For the Classical objects,  the fraction of total objects detected $f_{Classical}=0.027$ for $H \le 7.2$. The detected fraction of 3:2 objects is half as big, $f_{3:2} = 0.012$, while the fraction of Scattered objects detected is ten times smaller, with $f_{Scattered} = 0.0023$ (both for $H \le 7.5$). 

Note that the assumed discovery distance in the grid has an effect on the fraction calculated. If we instead assumed that objects were always found at perihelion, instead of at the median distance, the detected fractions for the Scattered and 3:2 objects are 8 times smaller. In the DES data, Classical objects were found, on average (median), close to the median distance ($+1.6$~AU from perihelion, and $-0.6$~AU from the median distance), while the 3:2 ($+3.5$ AU from perihelion and $-4.3$~AU from the median distance) and Scattered ($+2.6$ AU from perihelion and $-5.9$~AU from the median distance) objects were discovered somewhere in between. We chose to use the median distance rather than the perihelion (or the average distance of the discovered objects) to establish a minimum number of objects per class, and to avoid having the values of the grid depend on the properties of the objects found. 

The corresponding total number of objects (from maximizing Equation~\ref{eqn28} for $N_o'$) is found to be $N_{<7.2} = 4500 \pm 400$ Classical objects, $N_{<7.5} = 2500 \pm 450$ 3:2 objects, and $N_{<7.5} = 10000 \pm 2100$ Scattered objects (Table~\ref{table:mlfit}).  Compared to the CDF fit values (Table~\ref{table:numobj}), the ML method finds a larger total number of objects for all classes. At H=7.2, there are 10\% more Scattered, 30\% more Classical, and 40\% more 3:2 objects in the ML estimate than in the CDF estimate.

\section{Comparison to other observations}
\label{section:others}

\subsection{CFEPS}

Comparisons between the results of the two largest independent surveys for KBOs, CFEPS and DES, need to keep in mind three differences in observations and analysis.

\emph{1. Differences in observational strategies.} The specific strategies used have been described in detail elsewhere \citep[][]{Millis2002,Elliot2005, Petit2011, Gladman2012}, and may affect populations of observed and recovered objects. For instance, one should note that they used different filters ($g'$ for CFEPS and \emph{VR} for DES) and each cite results in $H$ magnitudes relative to those filters, althought the offset between them is small, roughly $g' - \text{\emph{VR}} = 0.1$ mag.  

One area of potential concern relates to the \emph{recovery bias}. The DES found a much larger number of objects initially, but due to limitations in telescope resources was unable to track and classify all of them, resulting in a smaller percentage recovered and classified compared to CFEPS. The breakdown of which types of DES objects were lost after initial discovery was reported in \citet{Elliot2005}, and lost objects tended to be fainter (requiring better follow-up conditions) and faster-moving (allowing for less time to recover an object before errors accumulated) than objects that were successfully tracked. Future work will explore how important the recovery bias is. We note that the advent of all-sky surveys like LSST and Pan-STARRS will mean that many of these objects will eventually be recovered and linked back to the original observations (with high quality orbits due to the long observational baselines), mitigating any current recovery bias.

\emph{2. Differences in class definitions.} Although mean-motion resonances are straightforward to identify, the precise meaning of a ``Classical" or ``Scattered" object can differ significantly between groups. The properties of smaller groups of objects can be particularly affect. The CFEPS Scattering population \citep[11 objects, ][]{Shankman2013} draws members from what the DES would classify as Scattered and Centaur groups, and as a result has a very different $H$ magnitude distribution than the DES Scattered class. Also, as noted earlier, differences in definitions result in high-inclination CFEPS Classical objects being classified as DES Scattered objects. Special care should be taken when comparing results between different groups and models to make sure the same objects are being analyzed.

\emph{3. Differences in statistical approach.} CFEPS follows a recursive backward-modeling approach, making synthetic populations of objects (drawn from distribution functions for $H$-$a$-$e$-$i$ as well as orbital longitude and libration amplitude for resonant objects). These synthetic populations are run through the CFEPS Survey Simulator to find which ones would have been detected. The properties of the initial distributions are adjusted until they match the recovered populations.

By contrast, the main DES approach is a forward-modeling method, where we calculate the biases of discovering each object at a randomized point in its orbit, and turn that into a discovery probability. Two methods are used to extrapolate from individual probabilities to class populations. The $H$-magnitude CDF method has each known object stand in for the (1 / probability) objects that we didn't find with similar parameters. The ML method fits for distributions of $H$-$a$-$e$-$i$, assuming functional forms such as exponentials or Lorentzians, for each class. Although the ML method results in larger absolute numbers, it produces results consistent with the CDF fits at the 1-2~$\sigma$ level. Given the inability to use the ML method for most dynamical classes, we use the CDF results to compare with the size distributions found by other groups.

\begin{figure}
\begin{centering}
\includegraphics*[width=0.7\textwidth]{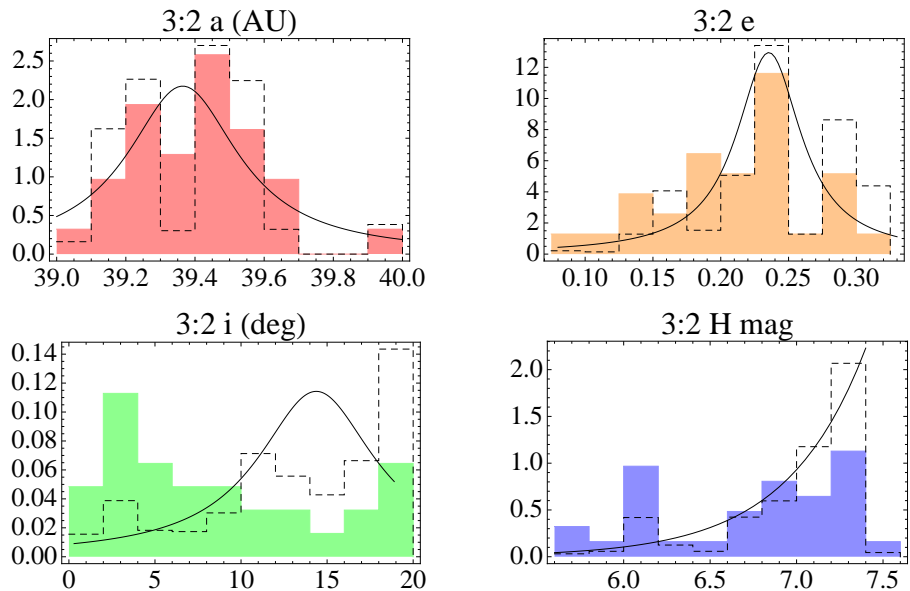}
\caption{Maximum likelihood fit to 32 3:2 objects with  $m_d \le 23$, and $4 \le H \le 7.5$. The filled histogram shows the observed distribution, while the dashed histogram shows the amount by which the distribution is adjusted to account for discovery probabilities. The black lines show the best maximum likelihood fit to each parameter.  
 }
\label{fig:ml3to2}
\end{centering}
\end{figure}


Despite the differences noted above, the CFEPS and DES results do agree on the \emph{slopes} for most classes. CFEPS has carved the main belt into several sections (``hot'', ``stirred'', and ``kernel''), with slopes of 0.8, 1.2, and 1.2 respectively \citep{Petit2011}; when combined together, as in Figure~\ref{fig:cdf}, the overall CFEPS distribution is similar in slope to the DES Classical slope $\alpha_{C}= 1.02\pm0.01$ (CDF method), with similar absolute numbers as well. (The ``hot'' and ``stirred''/``kernel'' slopes, meanwhile, are consistent with the results when the full DES sample is divided into ``Hot'' and ``Cold'' populations; see Section~\ref{section:fraser}). 

The \emph{absolute numbers} of CFEPS objects are generally the same order of magnitude as the DES results, but not entirely consistent, with CFEPS generally citing lower abundances. After accounting for the slight differences between results quoted in $H_g$ \citep{Petit2011} and \citet{Gladman2012} and $H_{VR}$ (this work), we find that the CFEPS Classical objects must be scaled upward by a factor of 1.7 to be consistent with the absolute number of DES objects (at the 5\% level using a KS test), while the 3:2 objects require a factor of 1.5. By contrast, the CFEPS Scattered objects are low by a factor of 14, although this discrepancy probably has more to do with the different definitions of ``Scattered"/``Scattering". The other, smaller resonant classes (2:1, 7:4, 5:2, and 5:3), which have few objects in either survey (and hence higher uncertainty) are all consistent without any rescaling.

The \emph{ratios between classes} also differ between the DES and CFEPS. Even ignoring the problematically-defined Scattered classes, the ratio of 3:2 to Classical objects is nearly 4 for the DES but around 6 for CFEPS. The ratio for 2:1 objects, meanwhile, is either a factor of 5 (DES) or 21 (CFEPS). The 5:2 abundances, on the other hand, are approximately the same in both surveys (7 times as many Classicals as 5:2s). Since resonance ratios are used to distinguish between planet formation scenarios, future work is need to resolve these discrepancies.

\subsection{Hot and Cold Classical populations}
\label{section:fraser}

There are many ways of subdividing populations of KBOs, and one long-standing division has been to divide the Classical belt between ``Hot" and ``Cold" objects, based on the orbital inclinations. We have noted before that the precise choice of class definition strongly impacts the values derived: for instance, several of the highest-inclination Classical objects in previous work \citep[e.g., ][]{Gulbis2010} have been classified here as Scattered objects, removing the need for a double Gaussian (or ``Hot Classical" population) inclination distribution. It is useful therefore to make comparisons wherever possible using the same classification scheme.

The definitions of ``Hot" and ``Cold" classes of \citet{Fraser2014} for \emph{all} objects (not just Classicals) have simple inclination and distance cuts as follows: $38 \le R \le 48$ and $i<5$ are ``Cold", while those from    $30 \le R \le 150$ and $5<i<90$ (``Hot"); and $30 \le R \le 38$ and $4<i<90$ (``Close"). We examined the DES sample using the same criteria (plus our restrictions to objects with $M\le23$ and $a\le80$~AU) and found that the best-fit slopes ($H\le7.2$) were $\alpha_{Hot} = 0.90 \pm 0.02$ and $\alpha_{Cold} = 1.32 \pm 0.02$. Despite using completely different objects and de-biasing methods, these numbers are very similar to Fraser's values: $\alpha_{Hot,Fraser} = 0.87^{+0.07}_{-0.2}$ with a break at $H=7.7^{+1.0}_{-0.5}$, and $\alpha_{Cold,Fraser} = 1.5^{+0.4}_{-0.2}$ with a break at $H=6.9^{+0.1}_{-0.2}$ \citep{Fraser2014}. The value after the break point for Fraser's Cold population (which is better constrained than the faint Hot population), is $\alpha_{2,Fraser} = 0.38^{+0.05}_{-0.09}$, entirely consistent with our Centaur slope estimate of  $\alpha = 0.42\pm0.02$.

\subsection{Rare and very distant objects}
\label{section:rare}

The lowest calculated probabilities fall into two general categories: objects with discovery magnitudes well below our average detection efficiency, and very distant objects. Seven of the ten most improbable objects found by the DES have $M_{disc} \ge 24$, and the remaining three have $a \ge 200$. We have already noted that our current analysis is restricted to objects brighter than $M_{disc}=23$ (future work using the magnitude recalibration may be able to extend that limit to fainter objects). We have also excluded the most distant objects ($a \le 80$~AU) after finding that a few objects widely scattered in semi-major-axis space leads to huge distortions in the distributions derived for all parameters. Only nine objects that otherwise met our selection criteria ($m_d \le 23$ and quality 3 orbits) were excluded, with $82\le a \le 653$~AU and classifications of Scattered-Near, Scattered-Extended, or Centaur. The CFEPS sample also has few objects in this range, with two scattering objects (2005 RH$_{52}$ and 2003 $HB_{57}$) around 150 AU \citep{Petit2011} and one (2003 $YQ_{179}$) reported in a 5:1 resonance at 88 AU.

The lowest detection probability for any object is $10^{-7}$ for 2001 FU$_{185}$, a very faint Scattered object ($M_{disc}=25.3$). The lowest probability with $M_{disc}\le23$ is $10^{-5}$ for 87269, an extremely eccentric Centaur with $a=653$~AU and $e=0.97$.  There are likely thousands of similarly far-flung objects, which are only detectable during a small fraction of their orbit, but characterizing their distribution parameters will require a larger sample and is beyond the scope of this work.

\begin{figure}
\begin{centering}
\includegraphics*[width=0.7\textwidth]{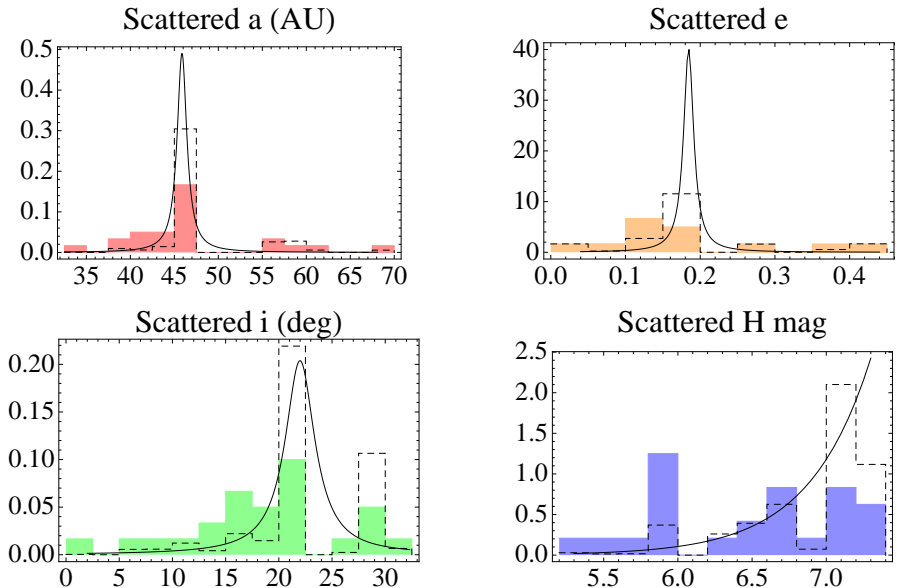}
\caption{Maximum likelihood fit to 24 Scattered objects with  $m_d \le 23$, and $H \le 7.5$. The filled histogram shows the observed distribution, while the dashed histogram shows the amount by which the distribution is adjusted to account for discovery probabilities. The black lines show the best maximum likelihood fit to each parameter.  
 }
\label{fig:mlscattered}
\end{centering}
\end{figure}

\section{Comparison to theory}
\label{section:theory}

As Neptune migrated outward during the early history of the solar system, it sculpted the primordial population of objects in the disk around and beyond its orbit, including pushing captured objects out into mean-motion resonances. In principle, we can use the relative number of objects in different dynamical classes to distinguish between models for how Neptune migrated, because objects in different orbits are captured with varying efficiency. Using the de-biased detection results of the previous sections, the DES finds, for instance, that there are about 0.7 times as many 2:1 objects as 3:2 objects (out to the break in the size distribution at $H_b=7.2$), and 1.3 times as many Scattered as Classical objects. We now examine recent models of planet formation and see how they compare to the DES class populations.

\subsection{Smooth migration}

The existence of objects in mean-motion resonances is evidence that Neptune migrated outward from where it formed \citep{Malhotra1993}. Many early numerical simulations \citep[e.g.,][]{Hahn2005} made the assumption that this outward migration happened more or less smoothly, with the giant planets maintaining generally low eccentricity, and that the population of objects that Neptune migrated through was dynamically ``cold'' (low $e$ and $i$). However, when such models attempted to reproduce the observed structure of the Kuiper Belt, they had difficulties explaining the observed eccentricity and inclination distributions. For instance, a pre-existing dynamically ``hot'' ($<e=0.1>$) disk allows for more efficient resonance capture into higher-order resonances, and \citet{Hahn2005} found that objects would be expected in many ``exotic'' resonances that the DES has in fact found objects in, including the 5:2, 9:4, 7:3, 11:6, 12:5, 3:1, 7:2, and 4:1 (Table~\ref{table:det}).

However, even with an initially excited disk, problems remain with smooth migration models. If the resonances were swept out of a hot Classical disk, the inclination distribution of the 3:2 resonance should resemble that of the Classical population it formed from, and this is not the case (compare Figures~\ref{fig:mlclassical} and \ref{fig:ml3to2}). Instead, the 3:2 inclination distribution is much closer to the Scattered inclination distribution (Figure~\ref{fig:mlscattered}), indicating a different formation model (see Section~\ref{section:chaos}). The smooth model is similarly argued against by the similarity in colors of the 3:2 and Scattered, but not Classical, populations \citep{Sheppard2012}.

Comparing absolute numbers of objects, we find that the smooth migration model of \citet{Hahn2005} overestimates the number of Classical objects by a factor of 7 (130,000 predicted $H\le9$, compared to the DES estimate of 19,000), and finds the opposite ratios of 3:2 and 2:1 objects (twice as many 2:1, whereas we find 0.7 times as many) and Classical to Scattered (five times as many Classical, whereas we find 30\% more Scattered). Other problems noted with smooth migration include the inability to explain objects with $i > 15$ degrees or to populate the extended scattered disk (perihelia beyond 40 AU), where objects such as Sedna and several DES objects have been found. For all of these reasons, smooth migration cannot fully explain the evolution of the outer solar system.

\subsection{Chaotic capture}
\label{section:chaos}

Current evidence suggests that the early history of the solar system was much more chaotic than previously thought, as proposed and later refined in a series of papers collectively referred to as the ``Nice Model''  \citep{Gomes2005, Tsiganis2005}. In its current incarnation \citep{Morbidelli2007, Levison2011, Nesvorny2012}, the model posits that a number of giant planets (up six) were originally configured so that Jupiter and Saturn were near a major resonance, likely a mutual 3:2 resonance, and dramatic changes in planetary orbits ensued when the resonance was crossed. In simulations with more than four giants, one or two planets similar to Uranus or Neptune were ejected from the early solar system; a major resonance crossing has also been invoked to explain the Late Heavy Bombardment of inner solar system. All of these events left a dramatic mark on the Kuiper Belt.

In general, the Nice Model has been much more successful than previous attempts at explaining Kuiper Belt structure. Regardless of precisely how objects are partitioned between Classical and Scattered designations, there exists a bi-modal inclination distribution \citep{Gulbis2010} which the Nice Model can reproduce. Additionally, the Nice Model predicted that some of the primordial population of objects was captured to make up Jupiter's Trojan population at the same time as the giant planets were sculpting the resonant and scattered populations of the Kuiper Belt \citep{Morbidelli2005, Nesvorny2013}, although observations at the time seemed to contradict the prediction that such objects would have similar slopes. A recent reanalysis of the Trojan size distribution by \citet{Fraser2014} has found that, in fact, the Trojan population has a similar two-slope fit to that work's results for the Hot population ($\alpha_{1,Trojan} =  1.0 \pm 0.2$, $\alpha_{2,Trojan} = 0.36 \pm 0.01$, and $H_{b,Trojan} = 8.4^{+0.2}_{-0.1}$), which are also in agreement with this work (except for some difference in the break point, which is likely due in part to $H$ magnitude conversions and differing average albedoes).

Nonetheless, some important details still need to be worked out. As noted in \citet{Levison2008}, the observed 3:2 distribution had higher eccentricities than the simulations could produce; as seen in Figure~\ref{fig:ml3to2}, the debiased eccentricity distribution is even more highly eccentric. The simulations also produced a Classical distribution that was too eccentric, although it has been noted by \citet{Batygin2011} that this may be less due to a failure of the model and more to the assumption that the space Neptune was pushing objects out into was previously unpopulated. \citet{Parker2010b} finds that a separate, primordial cold population is also supported by the high wide binary fraction, which is too large for the mechanism of \citet{Levison2008}.

Another interesting feature is that cold Classical objects (defined by inclination) are less eccentric than the stability limit requires \citep{Dawson2012}. For the least-excited ($i<2^{\circ}$) objects, there is a ``wedge" in $e-i$ space where no objects exist despite orbits being stable in that regime. \citet{Morbidelli2014} identified the area as just inside the 7:4 mean motion resonance. The wedge has been argued to support either a primordial cold belt of objects \citep{Dawson2012}, or to be a natural result of resonance interactions during smooth migration at the tail end of Neptune's outward journey \citep{Morbidelli2014}. Clear evidence of this wedge is seen in the DES Classical objects in the same region (between the 5:3 and 7:4 mean motion resonances), where there are 43 objects. We can do CDF and ML fits as above, and find that the CDF slope of the population just inside the 7:4 resonance ($42.3\le a\le 43.6$) agrees with the full Cold population, with $\alpha_{inside 7:4}=1.25\pm0.03$, which is not surprising given the ML fit to the inclinations ($i_o=1.7$, $i_s=1.7$), while the eccentricities are indeed well below the stability limit of $e=0.1$ ($e_o=0.043$, $e_s=0.012$). (We note that we find 10-15\% as many 7:4 as Classical objects in this region, or about half what the non-debiased results and model simulations of \citet{Morbidelli2014} both found; the difference is probably due to the slightly further distance, which adds 0.2 magnitudes, as well as to the higher inclinations and eccentricities of 7:4 objects, all of which slightly bias against discovery even of this relatively close class.) 

As more detailed model predictions of fine regions of phase space are made, it will become increasingly necessary that (a) all parties involved are using the same dynamical definitions, and (b) that observers make debiased data available and that modelers compare to debiased data wherever possible.

\subsection{Location of the power law break}
\label{section:break}

\begin{figure}
\begin{centering}
\includegraphics*[scale=0.7]{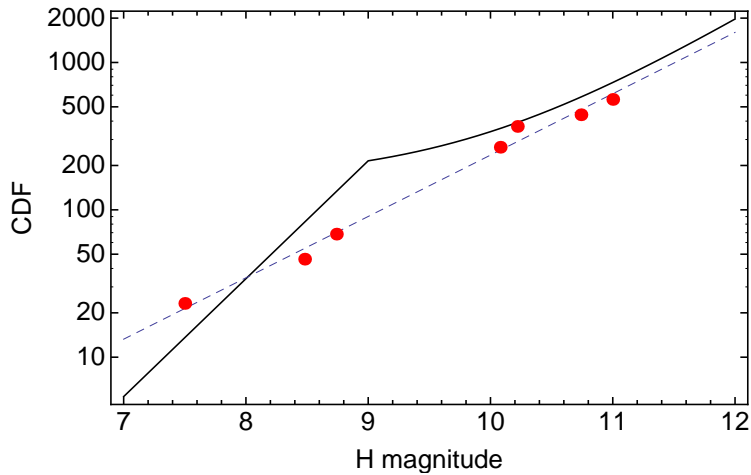}
\caption{$H$-magnitude distribution of Centaurs, which are well fit by a single power law with $\alpha_2 = 0.42\pm0.2$ (dashed line). No evidence is found for the divot reported by \citet{Shankman2013}, based on 11 CFEPS Scattered objects (solid line). (Note that the divot model has been scaled down by a factor of 20 from the absolute scale reported in \citet{Petit2011}, to match the values plotted in \citet{Shankman2013}.)  }
\label{fig:divot}
\end{centering}
\end{figure}

There has long been evidence that the size distribution of Kuiper Belt objects cannot follow a single power law. A break in the size distribution power law has been proposed for objects smaller than about 100 km diameter \citep{Gladman2001,Bernstein2004,Kenyon2004}, although the exact location varies by model. Destructive collisions have been evoked to explain the existence of a break at roughly the same size \citep[e.g., 40 km][]{Pan2005}. The exactly location of the size break, however, has not yet been observed because of the faint magnitudes of objects in this size regime: 100 km with an albedo of 0.05 roughly corresponds to $H=9$. 

Collisional evolution modeling by \citet{Fraser2009} presented theoretical support for a depletion region from $D=20-40$ km, with reduced collision rates for larger objects ($D=50-100$ km), allowing the steep power law from the brighter Kuiper Belt to continue until that point. A ``divot'' is expected in the power law at around that point. Recently, there has been a claimed detection of that ``divot'' around $H=9$ \citep{Shankman2013}, based on 11 Scattering objects discovered and de-biased by CFEPS using their Survey Simulator. These objects range in brightness from $H=7.1-10$, with only two objects (L4k09 and L4v11) fainter than the supposed ``divot'' point ($H=9.5$ and $H=10.0$, respectively). The lack of additional objects is itself used as a constraint on the steepness of the slope. 

Collisional grinding has come under fire recently, however, since it would have disrupted the many wide binaries found in the outer solar system \citep{Parker2012, Fraser2014}.  \citet{Fraser2014} has compiled detections up to $H=9$ and finds a much shallower turn-over at $H\sim7$, in line with what we have found in this independent work. After accounting for the albedo distribution, that paper concluded that the turn-over diameter is at $D=136 \pm 8$~km. The picture that has emerged is of a cold population (roughly the DES Classical group) that formed in situ around 40 AU, while the hot objects (DES Scattered plus resonances) formed around 15-35 AU and were scattered outward. The cold belt cannot have endured order-of-magnitude collisional evolution, and must have always had a surface density similar to the current density, while that of the hot objects must be $10^5$ times greater. These five orders of magnitudes imply a growth time rate that is $10^5$ times slower in the outer solar system, and this creates intractable problems for object formation time scales and the development of a broken size distribution \citep{Fraser2014}. Another theoretical explanation for the existence of a break is required.

Locating the break observationally also is challenging. Most observations do not have numerous objects on both sides, although meta-analyses such as \citet{Fraser2014} do a good job of pulling together disparate observations. The only DES class with objects fainter than about $H=8.5$ is the Centaurs, which approach closer than Neptune and are consequently often much brighter at discovery. Although Centaurs are now located inside the inner edge of the Kuiper Belt, they are thought to be the transitional stage between the scattered disk and Jupiter family comets (JFCs). This is supported by both dynamical evolution calculations \citep{Levison1997, Tiscareno2003} and the similarity of the Centaur inclination distribution \citep{Gulbis2010} to both the Scattered objects and JFCs (and not to the main Kuiper Belt or resonant objects). 

The slope observed for the Centaur objects, $\alpha_2 = 0.42\pm0.02$, is consistent with other populations of small objects reported in the literature, such as the Jupiter Family Comets \citep[$\alpha = 0.49\pm0.05$,][]{Solontoi2012}. It is also the same power law as the the faint end of the Jupiter Trojans as measured by \citet{Solontoi2012} ($\alpha = 0.44\pm0.05$) and \citet{Fraser2014} ($\alpha_{2,Trojan} = 0.36 \pm 0.01$). With 7 DES objects between $H=7.5-11$ and $m_d \leq 23$, we see no sign of the purported ``divot'' of \citet{Shankman2013}, which most of our objects are fainter than (Figure~\ref{fig:divot}). No sign of turnover is seen out to $H=11$ ($D=40$~km assuming albedo=0.05).


\section{Conclusions}
\label{section:conclusions}

The Kuiper Belt contains a record of outer solar system history, and de-biased observations of different classes of objects are a powerful tool to understanding it. The DES is the largest uniform survey to weigh in on the dynamical statistics. Here we have presented an analysis of a dataset of 304 objects, as well as a new method for accounting for discovery biases.

We have calculated the detection probability for 304 objects with secure orbital classifications, accounting for biases related to the object semi-major axis, eccentricity, inclination, discovery magnitude, discovery distance, and absolute magnitude. Using these probabilities, we can estimate the size distribution for 246 objects in 8 classes with at least 5 objects each, using $H$-magnitude as an observable proxy for object diameter. A double power-law is required to explain the data, with the population of Classical objects used to derive the main power law slope, $\alpha_1$, and the intrinsically fainter Centaur population is used to derive the slope after the break point, $\alpha_2$.  This power law is consistent with all eight classes that have sufficient objects. The parameters of this power law are $\alpha_1 = 1.02\pm0.01$ for $ H \le 7.2$ and $\alpha_2 = 0.42\pm0.02$ for fainter objects. 

We note that this power law also appears to break down at the very brightest, and largest, objects ($H<5$), as is seen in the 3:2 distribution. This may be due to small number statistics in the objects detected, the stochastic nature of a few large-body collisions, or a change in the underlying mechanisms that govern the formation of large objects.

We can also estimate the total number of objects in the Kuiper Belt up to a particular size. Two different methods were used, a max-likelihood estimation that determined distribution functions for four quantities ($a, e, i, H$) and a CDF based directly on inverted probabilities that only determined the $H$ magnitude distribution. For comparison between classes, we prefer the CDF results, which were possible on 8 classes (there were too few objects to use the max-likelihood approach for the other classes). We note that the ratios between the classes are similar in the max-likelihood analysis, although the overall abundances of objects are 1.5-2 times larger. We adopt the CDF results, and find that for $H\le 7$ the number of objects in 8 classes: Classical ($2100\pm300$ objects), Scattered ($2800\pm 400$), 3:2 ($570\pm 80$), 2:1 ($400\pm 50$), 5:2 ($270\pm 40$), 7:4 ($69\pm 9$), 5:3 ($60\pm 8$), and Centaurs ($13\pm 5$). 

Finally, we can compare our data to other reported observations and model predictions. The absolute number and power law slope agree with results presented by the other major Kuiper Belt survey, CFEPS, for the Classical objects. Some additional classes of objects (such as the 5:3 and 7:4 resonances) also agree. Others, notably the 3:2 resonance, differ in absolute numbers by a factor of 4. We also find no evidence for the divot reported in the Scattered population by \citet{Shankman2013}. These discrepancies in the details illustrate the value of having two completely independent data sets and analyses methods to determine the number of objects in the Kuiper Belt. 

Dividing our sample according to the ``Hot" and ``Cold" criteria of \citet{Fraser2014}, we find excellent agreement in the reported breaks and slopes. The evidence of that work and this strongly points toward a brighter turn-over, around $H=7$, than previously theorized. We would also like to reiterate the importance of comparing like to like, both in terms of dynamical class definitions and in terms of comparing observations to model results. 

In anticipation of large all-sky surveys such as LSST and Pan-STARRS, we are making this de-biasing code available\footnote{https://github.com/elisabethadams/des-class-populations} so that the discovery probabilities of objects found by any survey with the same characterizations can be calculated. We anticipate this will be a useful complement to tools such as the CFEPS Survey Simulator for determining the structure of the Kuiper Belt.

\acknowledgements

The authors would like to thank Alessondro Morbidelli, Brett Gladman and Wes Fraser for helpful discussions and constructive criticism, which greatly improved this paper. Research was supported in part by NASA Grants NAG5-13380, NAG5-11058, and NNG06-GI23G to Lowell Observatory; NSF Grants AST0406493 and AST0707609 to MIT; the Alfred P. Sloan Foundation at UCB, and NASA Grant nos. NAG5-4495 and NAG5-12236 at UH.  DET acknowledges support from the American Astronomical Society in the form of a Small Research Grant and from the Space Telescope Science Institute under grant GO-9433.06.  Support for program GO-9433 was provided by NASA, through a grant from the Space Telescope Science Institute, which is operated by the Association of Universities for Research in Astronomy (AURA), Inc. under NASA contract NAS 5-26555.  NOAO distributes the IRAF program, used for some of the image processing in this paper, and NOAO maintains the observing facilities used in this investigation.  NOAO is operated by AURA, under cooperative agreement with the National Science Foundation.  A. Gulbis specifically acknowledges support by the National Research Foundation of South Africa. S. Benecchi acknowledges support through a Carnegie Fellowship at the Department of Terrestrial Magnetism.

\clearpage

\bibliographystyle{astron}


\clearpage
\appendix

\LongTables
\begin{deluxetable}{c c  l l l l l l l l  }
\tablewidth{0pt}
\tablecaption{Detection probability for DES objects with secure classifications\tablenotemark{a}}
\tablehead{
Object & Class\tablenotemark{b}
 &  $a$& $e$ & $i$ & $H$ & $m_d$ & $R_d$ & Prob.\tablenotemark{c} & Fit\tablenotemark{d} \\
 &    & (AU) & & (deg) &  & (VR) & (AU) &
}
\startdata

  \text{2001QR322} & \text{1:1} & 30.38 & 0.0297 & 1.3438 & 7.42 & 21.1 & 29.654 & 0.26 & \text{} \\
\hline
 \text{2001FQ185} & \text{2:1} & 47.471 & 0.2258 & 2.2422 & 7.01 & 23.2 & 36.985 & 0.00016 & \text{} \\
 \text{2003FE128} & \text{2:1} & 47.714 & 0.2487 & 3.2774 & 6.17 & 21.5 & 36.019 & 0.019 & \text{} \\
 \text{2001UP18} & \text{2:1} & 48.008 & 0.0702 & 0.43497 & 5.71 & 22.5 & 50.271 & 0.061 & \text{} \\
 \text{2000QL251} & \text{2:1} & 48.043 & 0.2192 & 4.7652 & 6.54 & 22.2 & 38.211 & 0.011 & \text{} \\
 \text{2002VD130} & \text{2:1+4:2I} & 48.096 & 0.3272 & 3.3745 & 7.11 & 21.9 & 33.445 & 0.018 & \text{} \\
 \text{2004TV357} & \text{2:1} & 48.128 & 0.2827 & 11.155 & 6.63 & 21.6 & 36.839 & 0.013 & \text{} \\
 \text{2004VK78} & \text{2:1} & 48.207 & 0.3358 & 0.7672 & 8.16 & 22.4 & 32.549 & 0.008 & \text{} \\
\hline
 136120 & \text{3:1} & 61.837 & 0.4756 & 21.383 & 7.9 & 22.3 & 32.838 & 0.00068 & \text{} \\
\hline
 119069 & \text{3:2} & 39.022 & 0.2404 & 1.8456 & 7.1 & 22.6 & 37.156 & 0.055 & \text{Y} \\
 169071 & \text{3:2} & 39.094 & 0.1888 & 7.1794 & 8. & 24.5 & 34.69 & 0.000029 & \text{} \\
 \text{2005GV210} & \text{3:2} & 39.143 & 0.1727 & 11.806 & 6.75 & 22.8 & 40.774 & 0.014 & \text{Y} \\
 \text{2002GE32} & \text{3:2} & 39.164 & 0.2294 & 16.259 & 7.13 & 22.9 & 40.06 & 0.0093 & \text{Y} \\
 \text{2002GF32} & \text{3:2} & 39.166 & 0.178 & 2.213 & 5.69 & 21. & 42.181 & 0.32 & \text{Y} \\
 \text{2005GE187} & \text{3:2} & 39.208 & 0.3233 & 18.837 & 7.3 & 22.1 & 31.849 & 0.0081 & \text{Y} \\
 \text{2002CE251} & \text{3:2} & 39.231 & 0.2656 & 10.039 & 8.38 & 22.6 & 29.956 & 0.0024 & \text{} \\
 \text{2003FL127} & \text{3:2} & 39.236 & 0.2374 & 4.8246 & 6.17 & 22.4 & 47.025 & 0.12 & \text{Y} \\
 \text{2002GW31} & \text{3:2} & 39.265 & 0.244 & 3.8494 & 6.94 & 22.6 & 40.07 & 0.064 & \text{Y} \\
 \text{2001KY76} & \text{3:2} & 39.277 & 0.237 & 2.4311 & 6.09 & 21.5 & 39.057 & 0.2 & \text{Y} \\
 \text{2005GF187} & \text{3:2} & 39.287 & 0.252 & 2.7169 & 7.84 & 22.6 & 31.131 & 0.013 & \text{} \\
 \text{2005GB187} & \text{3:2} & 39.288 & 0.2331 & 15.558 & 7.04 & 21.6 & 30.156 & 0.01 & \text{Y} \\
 \text{2002GY32} & \text{3:2} & 39.298 & 0.0863 & 3.0394 & 6.87 & 21.5 & 35.978 & 0.16 & \text{Y} \\
 \text{2004EH96} & \text{3:2} & 39.302 & 0.2781 & 4.3025 & 8.13 & 21.9 & 28.56 & 0.021 & \text{} \\
 \text{2002GL32} & \text{3:2} & 39.324 & 0.1229 & 7.1363 & 7.85 & 22.6 & 34.495 & 0.0052 & \text{} \\
 \text{2002GV32} & \text{3:2} & 39.335 & 0.1886 & 3.9293 & 7.4 & 21.3 & 32.459 & 0.1 & \text{Y} \\
 307463 & \text{3:2} & 39.339 & 0.2068 & 2.9165 & 6.1 & 21.4 & 43.01 & 0.22 & \text{Y} \\
 \text{2003FF128} & \text{3:2} & 39.355 & 0.2122 & 0.3827 & 6.71 & 21.6 & 33.082 & 0.075 & \text{Y} \\
 \text{2004EV95} & \text{3:2} & 39.359 & 0.1871 & 12.544 & 7.4 & 23.3 & 41.424 & 0.0058 & \text{} \\
 \text{2003QB91} & \text{3:2} & 39.385 & 0.1898 & 5.3942 & 6.1 & 21.7 & 44.671 & 0.15 & \text{Y} \\
 \text{2004EJ96} & \text{3:2} & 39.401 & 0.237 & 9.1549 & 7.86 & 23. & 33.743 & 0.0028 & \text{} \\
 \text{2005EZ300} & \text{3:2} & 39.411 & 0.237 & 10.708 & 7.28 & 22.4 & 35.426 & 0.011 & \text{Y} \\
 306792 & \text{3:2} & 39.413 & 0.1559 & 17.038 & 7.2 & 21.7 & 36.772 & 0.024 & \text{Y} \\
 28978 & \text{3:2} & 39.42 & 0.2458 & 18.278 & 3.3 & 19. & 43.344 & 0.058 & \text{} \\
 133067 & \text{3:2} & 39.422 & 0.2565 & 9.6145 & 6.8 & 21.5 & 32.127 & 0.028 & \text{Y} \\
 \text{2000CK105} & \text{3:2} & 39.435 & 0.2317 & 9.2204 & 6.1 & 22.7 & 48.436 & 0.052 & \text{Y} \\
 \text{2001KD77} & \text{3:2} & 39.437 & 0.1115 & 1.5235 & 5.62 & 20.5 & 35.27 & 0.27 & \text{Y} \\
 \text{2001KB77} & \text{3:2} & 39.44 & 0.2806 & 18.514 & 7.25 & 21.7 & 31.665 & 0.011 & \text{Y} \\
 \text{2003QX111} & \text{3:2} & 39.464 & 0.1308 & 8.8664 & 6.59 & 21.5 & 39.314 & 0.078 & \text{Y} \\
 \text{2003QH91} & \text{3:2} & 39.468 & 0.1483 & 5.2109 & 6.65 & 22. & 41.727 & 0.1 & \text{Y} \\
 \text{2001QF298} & \text{3:2} & 39.499 & 0.1069 & 21.777 & 4.73 & 20. & 42.579 & 0.05 & \text{Y} \\
 69990 & \text{3:2} & 39.501 & 0.1907 & 7.8294 & 8.3 & 22.4 & 32.646 & 0.0063 & \text{} \\
 91205 & \text{3:2} & 39.507 & 0.1356 & 11.591 & 8. & 22.7 & 35.236 & 0.0031 & \text{} \\
 \text{2001RX143} & \text{3:2} & 39.51 & 0.2938 & 18.89 & 6.11 & 22.1 & 40.504 & 0.02 & \text{Y} \\
 \text{1998WV31} & \text{3:2} & 39.523 & 0.2728 & 4.5887 & 8.28 & 22.4 & 32.916 & 0.025 & \text{} \\
 \text{2004VT75} & \text{3:2} & 39.54 & 0.2112 & 12.373 & 6.26 & 21.5 & 36.495 & 0.036 & \text{Y} \\
 \text{2002CW224} & \text{3:2} & 39.548 & 0.2431 & 6.0052 & 6.93 & 22.3 & 39.025 & 0.052 & \text{Y} \\
 \text{2003UT292} & \text{3:2} & 39.556 & 0.299 & 18.213 & 6.93 & 21.1 & 30.936 & 0.018 & \text{Y} \\
 \text{2001QH298} & \text{3:2} & 39.568 & 0.1089 & 5.4677 & 7.67 & 22. & 36.576 & 0.05 & \text{} \\
 \text{2003UV292} & \text{3:2} & 39.59 & 0.218 & 12.168 & 7.22 & 22.2 & 33.906 & 0.01 & \text{Y} \\
 \text{2003WA191} & \text{3:2} & 39.604 & 0.2404 & 4.5802 & 8.5 & 22.1 & 30.237 & 0.016 & \text{} \\
 \text{2001RU143} & \text{3:2} & 39.612 & 0.1444 & 7.236 & 5.98 & 22.2 & 43.831 & 0.076 & \text{Y} \\
 139775 & \text{3:2} & 39.616 & 0.1982 & 6.0159 & 7.2 & 20.1 & 32.11 & 0.15 & \text{Y} \\
 \text{1998WS31} & \text{3:2} & 39.661 & 0.2064 & 6.6648 & 8.25 & 22.6 & 31.51 & 0.0036 & \text{} \\
 \text{2004VZ75} & \text{3:2} & 39.694 & 0.1913 & 4.2463 & 7.3 & 22.6 & 42.14 & 0.062 & \text{Y} \\
 69986 & \text{3:2} & 39.707 & 0.2292 & 15.089 & 7.9 & 22.3 & 31.282 & 0.0034 & \text{} \\
 \text{2002VD138} & \text{3:2} & 39.746 & 0.1505 & 4.12 & 8.44 & 22.8 & 34.886 & 0.0069 & \text{} \\
 \text{1998WZ31} & \text{3:2} & 39.758 & 0.1725 & 13.549 & 8.11 & 22.7 & 33.022 & 0.0015 & \text{} \\
 \text{1998UR43} & \text{3:2} & 39.78 & 0.2262 & 7.68 & 8.34 & 22.4 & 31.935 & 0.0072 & \text{} \\
 \text{2002VX130} & \text{3:2} & 39.79 & 0.2285 & 2.8342 & 8.35 & 22.1 & 30.811 & 0.027 & \text{} \\
 119473 & \text{3:2} & 39.971 & 0.292 & 3.1909 & 7.1 & 22.4 & 32.506 & 0.023 & \text{Y} \\
\hline
 \text{2003LA7} & \text{4:1} & 75.796 & 0.5269 & 5.1198 & 6.13 & 21.8 & 45.58 & 0.018 & \text{} \\
\hline
 \text{2005ER318} & \text{4:3} & 36.495 & 0.1606 & 11.352 & 7.72 & 22.1 & 31.417 & 0.011 & \text{} \\
 \text{1998UU43} & \text{4:3} & 36.682 & 0.1289 & 10.709 & 6.79 & 22.8 & 37.784 & 0.012 & \text{} \\
 \text{2004TX357} & \text{4:3} & 36.853 & 0.2165 & 15.522 & 8.55 & 22.6 & 29.361 & 0.0014 & \text{} \\
\hline
 119068 & \text{5:2} & 54.856 & 0.3544 & 11.751 & 6.7 & 21.5 & 35.418 & 0.0074 & \text{} \\
 135571 & \text{5:2+5:2I} & 55.041 & 0.3489 & 14.012 & 7.2 & 22.3 & 38.189 & 0.0032 & \text{} \\
 \text{2002GP32} & \text{5:2} & 55.098 & 0.4189 & 0.73692 & 6.99 & 20.3 & 32.163 & 0.09 & \text{} \\
 \text{2004EG96} & \text{5:2} & 55.248 & 0.4198 & 16.569 & 8.01 & 23.1 & 32.222 & 0.00014 & \text{} \\
 38084 & \text{5:2} & 55.305 & 0.4113 & 12.703 & 7.4 & 22. & 35.355 & 0.0047 & \text{} \\
 69988 & \text{5:2} & 55.543 & 0.4291 & 9.191 & 7.4 & 22.6 & 38.614 & 0.0056 & \text{} \\
 \text{2004TT357} & \text{5:2} & 56.026 & 0.438 & 10.095 & 7.44 & 22.2 & 31.511 & 0.0024 & \text{} \\
\hline
 \text{2002GS32} & \text{5:3} & 42.043 & 0.1054 & 5.8249 & 7.45 & 22.2 & 37.642 & 0.021 & \text{} \\
 \text{2000QN251} & \text{5:3} & 42.419 & 0.1282 & 1.3872 & 7.2 & 22.4 & 37.006 & 0.017 & \text{} \\
 143751 & \text{5:3+5:3I} & 42.515 & 0.2583 & 7.1891 & 8.6 & 22.6 & 31.536 & 0.0026 & \text{} \\
 149349 & \text{5:3} & 42.57 & 0.2425 & 8.102 & 6.8 & 22. & 40.042 & 0.026 & \text{} \\
 \text{2002VV130} & \text{5:3} & 42.647 & 0.1749 & 2.4766 & 7.51 & 22.7 & 37.151 & 0.012 & \text{} \\
\hline
 127871 & \text{5:4} & 34.895 & 0.0862 & 1.0169 & 7. & 21.6 & 32.714 & 0.17 & \text{} \\
 \text{2002GW32} & \text{5:4} & 34.906 & 0.0775 & 7.926 & 7.18 & 21.6 & 37.603 & 0.1 & \text{} \\
 \text{2003QB92} & \text{5:4} & 34.956 & 0.0901 & 2.151 & 7.24 & 22.2 & 37.35 & 0.2 & \text{} \\
\hline
 160148 & \text{7:2} & 69.147 & 0.5034 & 14.555 & 7.8 & 23.3 & 39.194 & 0.0003 & \text{} \\
\hline
 95625 & \text{7:3} & 52.81 & 0.3732 & 13.462 & 7.4 & 21.2 & 33.289 & 0.0089 & \text{} \\
 183964 & \text{7:3} & 53.08 & 0.3831 & 9.7465 & 7.4 & 22.5 & 33.7 & 0.0018 & \text{} \\
 \text{2002CZ248} & \text{7:3} & 53.207 & 0.3919 & 4.355 & 8.27 & 22.9 & 35.314 & 0.0031 & \text{} \\
\hline
 \text{1999HG12} & \text{7:4} & 43.443 & 0.1537 & 1.4773 & 7.18 & 22.5 & 43.109 & 0.057 & \text{} \\
 135024 & \text{7:4} & 43.502 & 0.1123 & 1.7254 & 6.8 & 23. & 45.805 & 0.038 & \text{} \\
 119070 & \text{7:4} & 43.517 & 0.1723 & 3.2705 & 7. & 21.6 & 36.02 & 0.066 & \text{} \\
 119066 & \text{7:4} & 43.537 & 0.0788 & 5.8422 & 6.8 & 22.2 & 42.698 & 0.043 & \text{} \\
 118378 & \text{7:4} & 43.577 & 0.1111 & 3.5257 & 7.4 & 23.3 & 41.634 & 0.0031 & \text{} \\
 119956 & \text{7:4} & 43.768 & 0.1724 & 2.5205 & 6.1 & 21. & 42.004 & 0.23 & \text{} \\
 \text{2000OP67} & \text{7:4} & 43.787 & 0.1882 & 1.6823 & 7.27 & 22.5 & 38.776 & 0.034 & \text{} \\
 \text{2003QW111} & \text{7:4} & 43.879 & 0.1138 & 1.485 & 6.44 & 21.4 & 45.068 & 0.25 & \text{} \\
 \text{2003QX91} & \text{7:4+7:4I} & 43.892 & 0.2535 & 28.088 & 8.22 & 22.7 & 33.107 & 0.00047 & \text{} \\
 \text{2004PW107} & \text{7:4} & 43.964 & 0.1368 & 3.483 & 6.55 & 21.3 & 38.875 & 0.12 & \text{} \\
 \text{2001QE298} & \text{7:4} & 43.966 & 0.1603 & 3.946 & 6.8 & 21.7 & 37.71 & 0.067 & \text{} \\
\hline
 \text{2001KG76} & \text{9:4} & 51.34 & 0.3397 & 0.021179 & 6.43 & 21.9 & 43.869 & 0.19 & \text{} \\
 182397 & \text{9:4} & 51.756 & 0.233 & 15.954 & 5.9 & 22.5 & 46.134 & 0.0067 & \text{} \\
 42301 & \text{9:4} & 51.871 & 0.2825 & 2.262 & 4.2 & 20.5 & 48.744 & 0.31 & \text{} \\
\hline
 \text{2002GD32} & \text{9:5} & 44.349 & 0.1406 & 6.3612 & 5.95 & 21.5 & 50.194 & 0.15 & \text{} \\
 \text{2001KL76} & \text{9:5} & 44.361 & 0.1006 & 2.8444 & 6.47 & 23.1 & 48.389 & 0.042 & \text{} \\
\hline
 182294 & \text{11:6} & 44.812 & 0.1593 & 9.7814 & 6.6 & 22.3 & 38.541 & 0.0066 & \text{} \\
\hline
 119878 & \text{12:5} & 53.982 & 0.3477 & 16.948 & 6.1 & 21.1 & 35.903 & 0.011 & \text{} \\
\hline
 54598 & \text{CENTAURR} & 16.564 & 0.2007 & 21.525 & 7.5 & 19.7 & 19.486 & 0.043 & \text{Y} \\
 \text{2003QD112} & \text{CENTAURR} & 19.089 & 0.5809 & 14.73 & 10.74 & 20.4 & 11.458 & 0.014 & \text{Y} \\
 \text{2003QC112} & \text{CENTAURR} & 20.553 & 0.2731 & 17.552 & 8.48 & 21.7 & 25.776 & 0.043 & \text{Y} \\
 \text{2003UY292} & \text{CENTAURR} & 22.023 & 0.2776 & 8.1602 & 10.22 & 21.9 & 15.909 & 0.0098 & \text{Y} \\
 \text{2000CO104} & \text{CENTAURR} & 24.057 & 0.1461 & 3.751 & 10.08 & 22.8 & 20.646 & 0.0051 & \text{Y} \\
 119976 & \text{CENTAURR} & 24.103 & 0.3884 & 3.7289 & 11. & 22.2 & 14.998 & 0.0084 & \text{Y} \\
 \text{2002PQ152} & \text{CENTAURR} & 25.891 & 0.1904 & 10.448 & 8.74 & 20.8 & 21.095 & 0.045 & \text{Y} \\
 88269 & \text{CENTAURR} & 25.912 & 0.2369 & 4.4225 & 9.5 & 23.4 & 22.414 & 0.0024 & \text{} \\
 120181 & \text{CENTAURR} & 32.566 & 0.178 & 2.0828 & 7.3 & 20.6 & 27.142 & 0.19 & \text{Y} \\
 \text{2004TE282} & \text{CENTAURR} & 34.65 & 0.1596 & 19.694 & 8.25 & 22.6 & 32.246 & 0.0043 & \text{Y}\tablenotemark{b} \\
 87555 & \text{CENTAURR} & 34.985 & 0.5622 & 7.8098 & 8.3 & 20.5 & 18.116 & 0.018 & \text{Y}\tablenotemark{b} \\
 149560 & \text{CENTAURR} & 41.617 & 0.4764 & 33.54 & 8.1 & 20.5 & 24.372 & 0.0056 & \text{Y}\tablenotemark{b} \\
 \text{2005EB299} & \text{CENTAURR} & 51.99 & 0.5106 & 1.8756 & 8.13 & 22.3 & 25.874 & 0.0063 & \text{Y}\tablenotemark{b} \\
 127546 & \text{CENTAURR} & 66.796 & 0.6858 & 76.346 & 8. & 20.6 & 21.973 & 0.0009 & \text{Y}\tablenotemark{b} \\
 \text{2003FH129} & \text{CENTAURR} & 71.318 & 0.6134 & 18.374 & 8.27 & 23.1 & 33.931 & 0.0006 & \text{} \\
 \text{2005PU21} & \text{CENTAURR} & 180.63 & 0.838 & 6.5096 & 6.32 & 22. & 49.206 & 0.0041 & \text{} \\
 87269 & \text{CENTAURS} & 652.78 & 0.9682 & 19.004 & 9.2 & 22.2 & 21.795 & 0.000014 & \text{} \\
\hline
 135182 & \text{CLASSICAL} & 37.24 & 0.0184 & 3.0487 & 6.2 & 22. & 36.885 & 0.12 & \text{Y} \\
 \text{2003FD128} & \text{CLASSICAL} & 38.166 & 0.0202 & 4.5304 & 7.26 & 23.2 & 37.555 & 0.0047 & \text{} \\
 \text{2003QA92} & \text{CLASSICAL} & 38.331 & 0.0621 & 1.9449 & 6.65 & 20.7 & 37.83 & 0.27 & \text{Y} \\
 82157 & \text{CLASSICAL} & 38.506 & 0.0575 & 4.632 & 6.9 & 23.5 & 38.98 & 0.003 & \text{} \\
 \text{2003QQ91} & \text{CLASSICAL} & 38.968 & 0.0734 & 4.2853 & 7.47 & 22.1 & 39.344 & 0.088 & \text{} \\
 \text{1998WV24} & \text{CLASSICAL} & 39.201 & 0.0403 & 2.1148 & 7.14 & 22.5 & 38.224 & 0.042 & \text{Y} \\
 80806 & \text{CLASSICAL} & 42.187 & 0.0687 & 3.0332 & 6.7 & 21.9 & 41.658 & 0.13 & \text{Y} \\
 88268 & \text{CLASSICAL} & 42.362 & 0.0165 & 0.43869 & 6.3 & 21.6 & 42.252 & 0.16 & \text{Y} \\
 \text{2005JZ174} & \text{CLASSICAL} & 42.457 & 0.075 & 5.3707 & 6.52 & 21.6 & 42.492 & 0.13 & \text{Y} \\
 \text{2003FK127} & \text{CLASSICAL} & 42.492 & 0.0465 & 0.82101 & 7.18 & 22.7 & 40.793 & 0.018 & \text{Y} \\
 \text{2002CD251} & \text{CLASSICAL} & 42.547 & 0.0141 & 1.223 & 7.16 & 22.9 & 42.512 & 0.021 & \text{Y} \\
 \text{2003QG91} & \text{CLASSICAL} & 42.607 & 0.0205 & 2.1091 & 7.27 & 22.8 & 42.325 & 0.023 & \text{} \\
 \text{2002VT130} & \text{CLASSICAL} & 42.683 & 0.0318 & 2.3887 & 5.77 & 21.5 & 42.653 & 0.19 & \text{Y} \\
 \text{2003QN91} & \text{CLASSICAL} & 42.718 & 0.1122 & 4.2729 & 7.22 & 21.6 & 38.031 & 0.082 & \text{} \\
 \text{2001QB298} & \text{CLASSICAL} & 42.744 & 0.0967 & 3.3341 & 6.91 & 21.4 & 39.037 & 0.14 & \text{Y} \\
 69987 & \text{CLASSICAL} & 42.761 & 0.0245 & 1.0048 & 7.2 & 22.8 & 42.019 & 0.017 & \text{Y} \\
 \text{2005JP179} & \text{CLASSICAL} & 42.783 & 0.0265 & 1.3795 & 5.86 & 20.7 & 42.16 & 0.26 & \text{Y} \\
 \text{1999HV11} & \text{CLASSICAL} & 42.784 & 0.0287 & 2.8516 & 7.78 & 22.8 & 43.578 & 0.03 & \text{} \\
 88267 & \text{CLASSICAL} & 42.836 & 0.025 & 1.0938 & 7.2 & 22.5 & 42.944 & 0.058 & \text{Y} \\
 \text{2003QY90} & \text{CLASSICAL} & 42.847 & 0.0503 & 2.2327 & 6.48 & 21.7 & 45. & 0.22 & \text{Y} \\
 \text{2003UN284} & \text{CLASSICAL} & 42.871 & 0.009 & 2.6743 & 7.42 & 23.2 & 42.486 & 0.0062 & \text{} \\
 \text{1999HJ12} & \text{CLASSICAL} & 42.883 & 0.0467 & 3.2114 & 7.32 & 22.6 & 44.188 & 0.048 & \text{} \\
 134860 & \text{CLASSICAL} & 42.948 & 0.019 & 0.446 & 6.1 & 21.7 & 42.612 & 0.15 & \text{Y} \\
 \text{2004DM71} & \text{CLASSICAL} & 43.008 & 0.033 & 0.79791 & 7.16 & 22.5 & 43.853 & 0.065 & \text{Y} \\
 \text{2003UT291} & \text{CLASSICAL} & 43.028 & 0.0554 & 1.2279 & 6.33 & 22.5 & 45.068 & 0.1 & \text{Y} \\
 \text{2000CL105} & \text{CLASSICAL} & 43.03 & 0.0458 & 2.7209 & 6.34 & 22.2 & 44.999 & 0.12 & \text{Y} \\
 \text{2002CS154} & \text{CLASSICAL} & 43.062 & 0.0457 & 0.44797 & 7.18 & 22.5 & 42.237 & 0.033 & \text{Y} \\
 \text{2004UD10} & \text{CLASSICAL} & 43.095 & 0.0311 & 2.551 & 6.46 & 22.5 & 44.132 & 0.057 & \text{Y} \\
 \text{2001FK185} & \text{CLASSICAL} & 43.105 & 0.0366 & 1.0018 & 7.38 & 25.7 & 41.721 & \text{7.8$*10^{-7}$} & \text{} \\
 \text{2001QO297} & \text{CLASSICAL} & 43.11 & 0.0385 & 1.1344 & 5.81 & 22.4 & 43.528 & 0.084 & \text{Y} \\
 \text{2003QE91} & \text{CLASSICAL} & 43.131 & 0.0496 & 3.6187 & 7.07 & 22.7 & 41.222 & 0.0061 & \text{Y} \\
 \text{2000ON67} & \text{CLASSICAL} & 43.138 & 0.0266 & 2.151 & 6.03 & 22.6 & 44.274 & 0.056 & \text{Y} \\
 \text{2003QF91} & \text{CLASSICAL} & 43.138 & 0.0391 & 1.3088 & 7.17 & 22.4 & 41.882 & 0.055 & \text{Y} \\
 160256 & \text{CLASSICAL} & 43.14 & 0.0598 & 3.3646 & 6.3 & 21.8 & 45.366 & 0.18 & \text{Y} \\
 \text{2003QE112} & \text{CLASSICAL} & 43.163 & 0.0441 & 2.6719 & 6.5 & 21.6 & 44.863 & 0.2 & \text{Y} \\
 \text{2003QL91} & \text{CLASSICAL} & 43.165 & 0.013 & 1.7288 & 6.62 & 21.9 & 42.72 & 0.17 & \text{Y} \\
 \text{2003QZ111} & \text{CLASSICAL} & 43.18 & 0.0642 & 3.8618 & 6.76 & 22. & 40.415 & 0.066 & \text{Y} \\
 \text{2000OU69} & \text{CLASSICAL} & 43.222 & 0.0493 & 2.9953 & 6.21 & 22.1 & 41.132 & 0.074 & \text{Y} \\
 \text{2001RW143} & \text{CLASSICAL} & 43.282 & 0.0385 & 1.9022 & 7.14 & 22.9 & 41.622 & 0.013 & \text{Y} \\
 \text{2002VB131} & \text{CLASSICAL} & 43.339 & 0.0356 & 0.26848 & 6.25 & 22.6 & 44.679 & 0.044 & \text{Y} \\
 \text{2002CZ154} & \text{CLASSICAL} & 43.349 & 0.0577 & 9.3039 & 7.4 & 22.7 & 40.881 & 0.0056 & \text{} \\
 \text{2002PD155} & \text{CLASSICAL} & 43.398 & 0.0104 & 6.9551 & 6.79 & 22.8 & 43.061 & 0.011 & \text{Y} \\
 \text{2003QD91} & \text{CLASSICAL} & 43.4 & 0.0341 & 0.83966 & 6.84 & 22.2 & 41.993 & 0.069 & \text{Y} \\
 \text{1999HH12} & \text{CLASSICAL} & 43.41 & 0.022 & 2.8173 & 7.02 & 22.6 & 44.081 & 0.048 & \text{Y} \\
 \text{2005EM303} & \text{CLASSICAL} & 43.413 & 0.0173 & 4.1308 & 7.47 & 23.1 & 42.758 & 0.0067 & \text{} \\
 \text{1998WY24} & \text{CLASSICAL} & 43.414 & 0.044 & 0.39369 & 6.58 & 22.3 & 41.856 & 0.037 & \text{Y} \\
 119067 & \text{CLASSICAL} & 43.431 & 0.1902 & 6.4212 & 6.6 & 21.9 & 44.138 & 0.073 & \text{Y} \\
 129772 & \text{CLASSICAL} & 43.511 & 0.0309 & 1.8241 & 7.2 & 22.6 & 42.245 & 0.035 & \text{Y} \\
 \text{2003QY111} & \text{CLASSICAL} & 43.526 & 0.0387 & 1.3893 & 6.71 & 22.1 & 42.107 & 0.088 & \text{Y} \\
 \text{1998WX24} & \text{CLASSICAL} & 43.558 & 0.0359 & 0.97684 & 6.57 & 22.8 & 45.12 & 0.032 & \text{Y} \\
 \text{2004PX107} & \text{CLASSICAL} & 43.595 & 0.053 & 1.892 & 7.04 & 23. & 41.468 & 0.0079 & \text{Y} \\
 \text{2003FM127} & \text{CLASSICAL} & 43.607 & 0.0614 & 3.3609 & 6.77 & 22.8 & 43.243 & 0.022 & \text{Y} \\
 \text{2003QU90} & \text{CLASSICAL} & 43.645 & 0.0582 & 3.0919 & 7.15 & 22.4 & 43.072 & 0.055 & \text{Y} \\
 \text{2005EC318} & \text{CLASSICAL} & 43.725 & 0.0368 & 1.0643 & 6.24 & 22.4 & 44.309 & 0.078 & \text{Y} \\
 \text{2002CU154} & \text{CLASSICAL} & 43.741 & 0.0593 & 1.8545 & 6.74 & 22.6 & 41.174 & 0.025 & \text{Y} \\
 \text{2003QX90} & \text{CLASSICAL} & 43.833 & 0.0209 & 2.1281 & 6.68 & 21.8 & 44.685 & 0.18 & \text{Y} \\
 \text{2002PQ145} & \text{CLASSICAL} & 43.855 & 0.0475 & 2.9071 & 5.51 & 20.7 & 45.877 & 0.25 & \text{Y} \\
 \text{2005GX186} & \text{CLASSICAL} & 43.859 & 0.0154 & 3.3568 & 6.65 & 23.6 & 43.442 & 0.0016 & \text{} \\
 \text{2001FK193} & \text{CLASSICAL} & 43.876 & 0.0657 & 1.9738 & 6.86 & 23.8 & 42.598 & 0.00078 & \text{} \\
 275809 & \text{CLASSICAL} & 43.884 & 0.0809 & 0.32597 & 5.6 & 20.2 & 42.546 & 0.2 & \text{Y} \\
 \text{2003QF113} & \text{CLASSICAL} & 43.907 & 0.0302 & 3.0095 & 6.55 & 22.1 & 42.587 & 0.088 & \text{Y} \\
 53311 & \text{CLASSICAL} & 43.912 & 0.0607 & 1.3276 & 6.6 & 22.6 & 43.841 & 0.045 & \text{Y} \\
 \text{2004VU75} & \text{CLASSICAL} & 43.923 & 0.1328 & 2.4578 & 6.55 & 22.4 & 41.889 & 0.052 & \text{Y} \\
 \text{2000CF105} & \text{CLASSICAL} & 43.927 & 0.0391 & 1.2035 & 6.82 & 22.8 & 42.243 & 0.018 & \text{Y} \\
 \text{2000CN114} & \text{CLASSICAL} & 43.929 & 0.0447 & 0.386 & 7.19 & 22. & 44.084 & 0.091 & \text{Y} \\
 \text{2004OL12} & \text{CLASSICAL} & 43.938 & 0.0675 & 2.3867 & 6.12 & 22.3 & 43.624 & 0.084 & \text{Y} \\
 \text{2005GC187} & \text{CLASSICAL} & 43.938 & 0.1223 & 2.9635 & 7.02 & 22.7 & 38.674 & 0.0092 & \text{Y} \\
 \text{1999HS11} & \text{CLASSICAL} & 43.945 & 0.0174 & 1.0874 & 6.54 & 22.5 & 43.775 & 0.054 & \text{Y} \\
 \text{2001DD106} & \text{CLASSICAL} & 43.958 & 0.0947 & 0.82671 & 6.86 & 23. & 40.619 & 0.0054 & \text{Y} \\
 \text{2002CB225} & \text{CLASSICAL} & 43.963 & 0.0776 & 2.2267 & 7.19 & 22.6 & 42.019 & 0.029 & \text{Y} \\
 \text{2004EU95} & \text{CLASSICAL} & 43.983 & 0.0392 & 1.6911 & 6.71 & 22.8 & 42.26 & 0.017 & \text{Y} \\
 \text{2005EF298} & \text{CLASSICAL} & 43.984 & 0.0883 & 1.522 & 6.05 & 21.4 & 41.025 & 0.16 & \text{Y} \\
 307616 & \text{CLASSICAL} & 43.997 & 0.0777 & 10.156 & 5.4 & 19.9 & 44.485 & 0.087 & \text{Y} \\
 \text{2005EX297} & \text{CLASSICAL} & 44.006 & 0.1102 & 4.4748 & 6.11 & 23.1 & 45.317 & 0.015 & \text{} \\
 \text{2000CE105} & \text{CLASSICAL} & 44.011 & 0.0597 & 1.0616 & 6.82 & 23. & 41.383 & 0.0063 & \text{Y} \\
 \text{2001FL185} & \text{CLASSICAL} & 44.039 & 0.0745 & 2.1814 & 7.12 & 23.5 & 40.889 & 0.0012 & \text{} \\
 \text{2003QV90} & \text{CLASSICAL} & 44.061 & 0.0551 & 1.6099 & 6.95 & 22.1 & 43.899 & 0.11 & \text{Y} \\
 \text{2002XH91} & \text{CLASSICAL} & 44.067 & 0.0851 & 3.5128 & 5.52 & 21.5 & 47.69 & 0.21 & \text{Y} \\
 \text{2002FX36} & \text{CLASSICAL} & 44.095 & 0.0516 & 2.6337 & 6.28 & 22.2 & 45.661 & 0.11 & \text{Y} \\
 \text{2000CJ105} & \text{CLASSICAL} & 44.108 & 0.1095 & 10.843 & 5.86 & 21.9 & 47.533 & 0.054 & \text{Y} \\
 88611 & \text{CLASSICAL} & 44.131 & 0.0243 & 4.0269 & 5.8 & 21.8 & 44.967 & 0.14 & \text{Y} \\
 \text{2004PY107} & \text{CLASSICAL} & 44.169 & 0.0947 & 0.41522 & 6.24 & 21.4 & 41.055 & 0.13 & \text{Y} \\
 \text{2002PO149} & \text{CLASSICAL} & 44.173 & 0.0598 & 0.99542 & 6.46 & 21.6 & 46.098 & 0.2 & \text{Y} \\
 \text{2001QQ322} & \text{CLASSICAL} & 44.191 & 0.0557 & 2.5497 & 6.23 & 21.6 & 43.626 & 0.18 & \text{Y} \\
 \text{2002CY154} & \text{CLASSICAL} & 44.196 & 0.077 & 0.77915 & 6.47 & 22.9 & 47.593 & 0.05 & \text{Y} \\
 \text{2001QJ298} & \text{CLASSICAL} & 44.214 & 0.0409 & 2.0786 & 6.14 & 21.4 & 45.232 & 0.22 & \text{Y} \\
 \text{2000CP104} & \text{CLASSICAL} & 44.215 & 0.0993 & 8.238 & 6.66 & 22.5 & 46.784 & 0.033 & \text{Y} \\
 60454 & \text{CLASSICAL} & 44.227 & 0.0832 & 2.5344 & 6.3 & 22. & 44.004 & 0.12 & \text{Y} \\
 \text{2001QS322} & \text{CLASSICAL} & 44.231 & 0.0428 & 1.6546 & 5.92 & 21. & 42.342 & 0.26 & \text{Y} \\
 \text{2001KF76} & \text{CLASSICAL} & 44.25 & 0.0255 & 2.4047 & 7.11 & 22.7 & 44.466 & 0.037 & \text{Y} \\
 183963 & \text{CLASSICAL} & 44.263 & 0.0996 & 0.91072 & 7.2 & 22.8 & 42.219 & 0.013 & \text{Y} \\
 \text{2000OH67} & \text{CLASSICAL} & 44.266 & 0.0184 & 4.1841 & 6.47 & 22.4 & 43.561 & 0.045 & \text{Y} \\
 \text{2001QX297} & \text{CLASSICAL} & 44.269 & 0.031 & 2.4712 & 6.38 & 22.1 & 43.615 & 0.11 & \text{Y} \\
 \text{2003QB112} & \text{CLASSICAL} & 44.3 & 0.1111 & 11.241 & 7.43 & 21.8 & 39.719 & 0.023 & \text{} \\
 \text{2004DN64} & \text{CLASSICAL} & 44.316 & 0.0521 & 1.0277 & 7.58 & 23.5 & 42.024 & 0.0012 & \text{} \\
 \text{2001RZ143} & \text{CLASSICAL} & 44.343 & 0.0695 & 2.5973 & 6.01 & 22.2 & 41.463 & 0.051 & \text{Y} \\
 \text{2001QZ297} & \text{CLASSICAL} & 44.351 & 0.0636 & 2.2219 & 6.98 & 22.6 & 41.863 & 0.025 & \text{Y} \\
 \text{2002VF131} & \text{CLASSICAL} & 44.361 & 0.0425 & 1.7053 & 6.63 & 22.5 & 44.5 & 0.064 & \text{Y} \\
 \text{2000CL104} & \text{CLASSICAL} & 44.387 & 0.0786 & 1.0955 & 6.1 & 22. & 42.564 & 0.11 & \text{Y} \\
 \text{2003QJ91} & \text{CLASSICAL} & 44.419 & 0.054 & 1.0266 & 6.48 & 22.5 & 44.925 & 0.055 & \text{Y} \\
 \text{2003QT91} & \text{CLASSICAL} & 44.431 & 0.0902 & 1.4422 & 6.67 & 22.1 & 40.913 & 0.056 & \text{Y} \\
 \text{2001UN18} & \text{CLASSICAL} & 44.494 & 0.0697 & 3.5001 & 6.46 & 22.3 & 46.351 & 0.096 & \text{Y} \\
 148780 & \text{CLASSICAL} & 44.495 & 0.0597 & 5.5508 & 5.7 & 22.2 & 45.101 & 0.071 & \text{Y} \\
 \text{2001QR297} & \text{CLASSICAL} & 44.504 & 0.0291 & 4.3395 & 6.28 & 21.6 & 43.633 & 0.14 & \text{Y} \\
 \text{2003QA91} & \text{CLASSICAL} & 44.505 & 0.0706 & 0.89682 & 5.5 & 21. & 45.262 & 0.22 & \text{Y} \\
 \text{2000CN105} & \text{CLASSICAL} & 44.506 & 0.0995 & 3.1799 & 4.89 & 21.4 & 45.692 & 0.19 & \text{Y} \\
 \text{2001QQ297} & \text{CLASSICAL} & 44.6 & 0.0862 & 2.9425 & 6.68 & 22. & 42.362 & 0.094 & \text{Y} \\
 \text{2005EN302} & \text{CLASSICAL} & 44.687 & 0.0665 & 2.3497 & 6.97 & 23.3 & 47.443 & 0.015 & \text{} \\
 \text{2004DM64} & \text{CLASSICAL} & 44.797 & 0.1296 & 3.3272 & 6.78 & 22.8 & 41.55 & 0.013 & \text{Y} \\
 \text{1998WW31} & \text{CLASSICAL} & 44.906 & 0.0828 & 8.0429 & 6.16 & 22. & 46.577 & 0.06 & \text{Y} \\
 \text{2001KT76} & \text{CLASSICAL} & 44.929 & 0.0855 & 2.273 & 6.92 & 22.9 & 42.3 & 0.012 & \text{Y} \\
 184314 & \text{CLASSICAL} & 44.98 & 0.1375 & 6.1414 & 6.3 & 22. & 42.409 & 0.05 & \text{Y} \\
 \text{2002CZ224} & \text{CLASSICAL} & 45.002 & 0.063 & 0.85619 & 7.14 & 23. & 47.627 & 0.026 & \text{Y} \\
 \text{2004ES95} & \text{CLASSICAL} & 45.104 & 0.1355 & 3.0388 & 6.77 & 22.5 & 41.197 & 0.025 & \text{Y} \\
 \text{2003UN292} & \text{CLASSICAL} & 45.221 & 0.1139 & 1.5537 & 7.34 & 23.3 & 41.13 & 0.0017 & \text{} \\
 160091 & \text{CLASSICAL} & 45.264 & 0.1089 & 3.2665 & 6.8 & 22.7 & 42.269 & 0.016 & \text{Y} \\
 \text{2001OQ108} & \text{CLASSICAL} & 45.36 & 0.0125 & 1.3674 & 6.53 & 22.9 & 45.757 & 0.02 & \text{Y} \\
 149348 & \text{CLASSICAL} & 45.365 & 0.1252 & 4.5418 & 6.3 & 21.8 & 41.269 & 0.072 & \text{Y} \\
 \text{2002VD131} & \text{CLASSICAL} & 45.384 & 0.0649 & 0.74774 & 6.58 & 21.9 & 43.304 & 0.14 & \text{Y} \\
 \text{2003UZ291} & \text{CLASSICAL} & 45.399 & 0.1306 & 6.4194 & 6.8 & 22.7 & 48.022 & 0.034 & \text{Y} \\
 \text{2003LB7} & \text{CLASSICAL} & 45.403 & 0.1282 & 1.682 & 6.68 & 21.4 & 40.089 & 0.15 & \text{Y} \\
 \text{2003GF55} & \text{CLASSICAL} & 45.407 & 0.0616 & 5.5284 & 6.51 & 23.3 & 46.455 & 0.0056 & \text{} \\
 \text{2001QP297} & \text{CLASSICAL} & 45.409 & 0.1177 & 0.41271 & 6.42 & 21.8 & 43.304 & 0.088 & \text{Y} \\
 \text{1998WY31} & \text{CLASSICAL} & 45.43 & 0.1178 & 1.0825 & 6.96 & 22.8 & 45.303 & 0.032 & \text{Y} \\
 \text{2005EO304} & \text{CLASSICAL} & 45.479 & 0.065 & 1.8596 & 6.28 & 22.5 & 43.622 & 0.041 & \text{Y} \\
 \text{2003QO91} & \text{CLASSICAL} & 45.555 & 0.1356 & 6.0962 & 6.71 & 21.4 & 39.457 & 0.068 & \text{Y} \\
 \text{1998WX31} & \text{CLASSICAL} & 45.707 & 0.1123 & 2.5532 & 6.59 & 22.3 & 40.66 & 0.028 & \text{Y} \\
 \text{1998WG24} & \text{CLASSICAL} & 45.828 & 0.1308 & 0.71848 & 6.51 & 22. & 41.318 & 0.083 & \text{Y} \\
 \text{2001KH76} & \text{CLASSICAL} & 46.018 & 0.123 & 4.7767 & 6.47 & 22. & 44.767 & 0.078 & \text{Y} \\
 19521 & \text{CLASSICAL} & 46.027 & 0.1081 & 11.003 & 4.8 & 20.2 & 42.464 & 0.076 & \text{Y} \\
 \text{2000CQ114} & \text{CLASSICAL} & 46.051 & 0.1155 & 2.2981 & 6.59 & 22.4 & 45.05 & 0.066 & \text{Y} \\
 \text{2002CY248} & \text{CLASSICAL} & 46.245 & 0.1435 & 8.4779 & 5.17 & 21.4 & 52.057 & 0.094 & \text{Y} \\
 \text{2001FO185} & \text{CLASSICAL} & 46.298 & 0.1162 & 11.267 & 6.74 & 24.2 & 41.188 & 0.000021 & \text{} \\
 \text{2003QS91} & \text{CLASSICAL} & 46.336 & 0.1449 & 4.3372 & 7.4 & 22.2 & 40.207 & 0.024 & \text{} \\
 \text{2004DH64} & \text{CLASSICAL} & 46.456 & 0.0913 & 2.4868 & 6.08 & 22.2 & 50.066 & 0.15 & \text{Y} \\
 \text{2003QR91} & \text{CLASSICAL} & 46.641 & 0.1828 & 5.0561 & 6.31 & 20.8 & 38.678 & 0.11 & \text{Y} \\
 \text{2002CT154} & \text{CLASSICAL} & 46.702 & 0.1116 & 3.1504 & 7.14 & 22.8 & 41.766 & 0.0069 & \text{Y} \\
 \text{2005JR179} & \text{CLASSICAL} & 46.704 & 0.1149 & 2.4388 & 5.83 & 22. & 45.303 & 0.11 & \text{Y} \\
 138537 & \text{CLASSICAL} & 46.734 & 0.1437 & 5.1841 & 5.9 & 21.7 & 40.895 & 0.057 & \text{Y} \\
 126719 & \text{CLASSICAL} & 46.906 & 0.1898 & 2.3308 & 6.7 & 22.1 & 38.864 & 0.036 & \text{Y} \\
 \text{2003LD9} & \text{CLASSICAL} & 47.071 & 0.1698 & 8.2553 & 6.57 & 22. & 40.601 & 0.019 & \text{Y} \\
 \text{2001KA77} & \text{CLASSICAL} & 47.289 & 0.0957 & 13.216 & 5.12 & 21.2 & 48.945 & 0.057 & \text{Y} \\
 \text{2003UY291} & \text{CLASSICAL} & 49.592 & 0.1669 & 4.0743 & 7.18 & 22.8 & 42.758 & 0.0064 & \text{Y} \\
 182933 & \text{CLASSICAL} & 50.24 & 0.2382 & 0.68281 & 6.4 & 21.4 & 42.354 & 0.14 & \text{Y} \\
\hline
 38083 & \text{SCATEXTD} & 38.89 & 0.1559 & 12.775 & 6.7 & 22.1 & 38.252 & 0.024 & \text{Y} \\
 \text{2004PA112} & \text{SCATEXTD} & 39.093 & 0.1124 & 32.223 & 7.06 & 21.7 & 35.836 & 0.011 & \text{Y} \\
 \text{2004PT107} & \text{SCATEXTD} & 40.643 & 0.0604 & 27.264 & 5.98 & 20.8 & 38.438 & 0.03 & \text{Y} \\
 \text{2001FU185} & \text{SCATEXTD} & 41.406 & 0.1655 & 23.798 & 9. & 25.3 & 34.555 & \text{1.5$*10^{-7}$} & \text{} \\
 \text{2002GH32} & \text{SCATEXTD} & 41.885 & 0.0914 & 27.646 & 5.37 & 20.8 & 42.556 & 0.032 & \text{Y} \\
 \text{2005JA175} & \text{SCATEXTD} & 42.293 & 0.1098 & 15.32 & 5.72 & 21. & 46.617 & 0.054 & \text{Y} \\
 \text{2001FN185} & \text{SCATEXTD} & 42.369 & 0.0695 & 22.27 & 7.11 & 24.5 & 39.524 & \text{5.1$*10^{-6}$} & \text{}
\\
 \text{2003QA112} & \text{SCATEXTD} & 42.795 & 0.1142 & 16.223 & 5.95 & 21.4 & 47.197 & 0.048 & \text{Y} \\
 \text{2004DG77} & \text{SCATEXTD} & 43.689 & 0.1244 & 48.138 & 7.14 & 23.7 & 45.831 & 0.00031 & \text{} \\
 \text{2001KO77} & \text{SCATEXTD} & 43.767 & 0.1434 & 22.118 & 7.68 & 22.4 & 37.887 & 0.0026 & \text{} \\
 182934 & \text{SCATEXTD} & 44.086 & 0.1092 & 12.989 & 5.4 & 20.5 & 42.751 & 0.064 & \text{Y} \\
 138628 & \text{SCATNEAR} & 44.997 & 0.2687 & 16.065 & 7.1 & 22. & 35.027 & 0.0063 & \text{Y} \\
 118379 & \text{SCATEXTD} & 45.131 & 0.2333 & 14.211 & 7.6 & 22.3 & 39.156 & 0.0081 & \text{} \\
 \text{2004PB108} & \text{SCATEXTD} & 45.292 & 0.112 & 19.243 & 6.45 & 21.9 & 43.72 & 0.018 & \text{Y} \\
 \text{2001KE77} & \text{SCATEXTD} & 45.311 & 0.1823 & 22.3 & 7.19 & 23. & 38.758 & 0.00059 & \text{Y} \\
 \text{2004PZ107} & \text{SCATEXTD} & 45.721 & 0.1876 & 12.344 & 7.3 & 22.3 & 38.658 & 0.006 & \text{Y} \\
 \text{2001KW76} & \text{SCATEXTD} & 45.869 & 0.2149 & 9.2413 & 7.7 & 22.9 & 40.328 & 0.0036 & \text{} \\
 \text{2002VF130} & \text{SCATEXTD} & 46.075 & 0.1197 & 20.967 & 7.15 & 22.6 & 42.368 & 0.0029 & \text{Y} \\
 \text{2001QA298} & \text{SCATEXTD} & 46.301 & 0.1902 & 22.2 & 6.68 & 22. & 38.643 & 0.0054 & \text{Y} \\
 181855 & \text{SCATEXTD} & 46.31 & 0.1879 & 27.769 & 7.2 & 22.7 & 38.915 & 0.00095 & \text{Y} \\
 \text{2000CG105} & \text{SCATEXTD} & 46.358 & 0.0417 & 29.237 & 6.47 & 22.8 & 46.423 & 0.0027 & \text{Y} \\
 \text{2002CX224} & \text{SCATEXTD} & 46.437 & 0.1318 & 15.968 & 5.99 & 22.8 & 48.728 & 0.0095 & \text{Y} \\
 \text{2001FT185} & \text{SCATEXTD} & 46.928 & 0.1002 & 19.326 & 7.74 & 24.6 & 43.043 & \text{4.$*10^{-6}$} & \text{} \\
 \text{2000CO105} & \text{SCATEXTD} & 47.157 & 0.1482 & 20.614 & 5.92 & 22.5 & 49.289 & 0.01 & \text{Y} \\
 \text{2003UB292} & \text{SCATEXTD} & 47.359 & 0.0493 & 18.719 & 5.97 & 22.5 & 49.673 & 0.013 & \text{Y} \\
 \text{2004PS107} & \text{SCATEXTD} & 47.94 & 0.2354 & 23.408 & 7.59 & 23.4 & 37.297 & 0.000092 & \text{} \\
 \text{2004OL14} & \text{SCATNEAR} & 48.878 & 0.3063 & 19.849 & 8.08 & 22.4 & 34.016 & 0.0013 & \text{} \\
 181874 & \text{SCATEXTD} & 52.492 & 0.2554 & 17.616 & 6.8 & 24.7 & 41.392 & \text{2.5$*10^{-6}$} & \text{} \\
 \text{2004OJ14} & \text{SCATEXTD} & 55.245 & 0.2887 & 20.944 & 6.37 & 22.4 & 44.224 & 0.0035 & \text{Y} \\
 \text{2000CQ105} & \text{SCATEXTD} & 56.917 & 0.3909 & 18.405 & 5.93 & 21.8 & 50.912 & 0.014 & \text{Y} \\
 60458 & \text{SCATEXTD} & 59.395 & 0.4022 & 20.983 & 6.6 & 22.7 & 44.109 & 0.0026 & \text{Y} \\
 \text{2001KG77} & \text{SCATNEAR} & 61.431 & 0.4474 & 16.918 & 8.06 & 23.1 & 34.939 & 0.00017 & \text{} \\
 134210 & \text{SCATEXTD} & 62.435 & 0.3972 & 7.7491 & 6.8 & 21.4 & 37.851 & 0.013 & \text{Y} \\
 \text{2003QK91} & \text{SCATEXTD} & 68.016 & 0.4347 & 5.5719 & 6.63 & 21.8 & 40.502 & 0.013 & \text{Y} \\
 \text{2004TF282} & \text{SCATNEAR} & 81.825 & 0.5185 & 24.326 & 6.08 & 21.9 & 39.431 & 0.0011 & \text{} \\
 \text{2005EF304} & \text{SCATEXTD} & 85.899 & 0.5514 & 18.55 & 7.24 & 22.8 & 40.06 & 0.00028 & \text{} \\
 118702 & \text{SCATEXTD} & 99.268 & 0.6044 & 24.375 & 6.8 & 21.9 & 40.231 & 0.00093 & \text{} \\
 184212 & \text{SCATNEAR} & 109.54 & 0.6773 & 15.763 & 7.2 & 21.9 & 35.896 & 0.0011 & \text{} \\
 \text{2002GB32} & \text{SCATNEAR} & 203.1 & 0.826 & 14.025 & 7.61 & 21.7 & 37.146 & 0.00085 & \text{} \\
 82158 & \text{SCATNEAR} & 213.16 & 0.8394 & 30.665 & 6. & 22.3 & 34.342 & 0.000072 & \text{} \\
 148209 & \text{SCATNEAR} & 225.59 & 0.8047 & 21.435 & 6.3 & 22.6 & 52.712 & 0.00033 & \text{} \\
 \text{2003FH127} & \text{SCATNEAR} & 738.63 & 0.95 & 1.262 & 7.23 & 23.3 & 36.978 & \text{4.$*10^{-6}$} & \text{} \\

 \enddata
\tablenotetext{a}{Full version of Table~\ref{table:prob}.}
\tablenotetext{b}{Centaurs with $a_{Nep}=30.1<a<80$~AU were grouped with Scattered objects for fits and plots throughout this paper. }
\tablenotetext{c}{Probability of detecting object with listed parameters and randomized ecliptic longitude.}
\tablenotetext{d}{Object used to derive fits. Centaur and Classical objects were used to derive CDF fits, and Classical, 3:2 and Scattered classes were used for maximum likelihood fits.}

\label{table:probfull}
\end{deluxetable}



\end{document}